\documentclass[superscriptaddress,twocolumn,pre]{revtex4}

\usepackage{amsmath}
\usepackage{graphicx,color}

\newcommand{\be}{\begin{equation}}
\newcommand{\ee}{\end{equation}}
\newcommand{\bea}{\begin{eqnarray}}
\newcommand{\eea}{\end{eqnarray}}

\begin{document}
\sloppy


\title{Instability of a uniformly collapsing cloud of classical and \\
quantum self-gravitating Brownian particles}

\author{Pierre-Henri Chavanis}
\affiliation{Laboratoire de Physique Th\'eorique (IRSAMC), CNRS and UPS, Universit\'e de Toulouse, France}

\begin{abstract}
We study the growth of perturbations in a uniformly collapsing cloud of self-gravitating Brownian particles. This problem shares analogies with the formation of large-scale structures in a universe experiencing a ``big-crunch'' or with the formation of stars in a molecular cloud experiencing  gravitational collapse. Starting from the barotropic Smoluchowski-Poisson system, we derive a new equation describing the evolution of the density contrast in the comoving (collapsing) frame. This equation can serve as a prototype to study the process of self-organization in complex media with structureless initial conditions. We solve this equation analytically in the linear regime and compare the results with those obtained by using the ``Jeans swindle'' in a static medium. The stability criteria, as well as the laws for the time evolution of  the perturbations, are different. The Jeans criterion is expressed in terms of a critical wavelength $\lambda_J$ while our criterion is expressed in terms of a critical polytropic index $\gamma_{4/3}$. In a static background, the system is stable for $\lambda<\lambda_J$ and unstable for $\lambda>\lambda_J$. In a collapsing cloud, the system is stable for $\gamma>\gamma_{4/3}$ and unstable for $\gamma<\gamma_{4/3}$. If $\gamma=\gamma_{4/3}$, it is stable for $\lambda<\lambda_J$ and unstable for $\lambda>\lambda_J$. We also study the fragmentation process in the nonlinear regime. We determine the growth of the skewness, the long-wavelength tail of the power spectrum and find a self-similar solution to the nonlinear equations valid for large times. Finally, we consider dissipative self-gravitating Bose-Einstein condensates with short-range interactions and show that, in a strong friction limit, the dissipative Gross-Pitaevskii-Poisson system is equivalent to the quantum barotropic  Smoluchowski-Poisson system. This yields a new type of nonlinear mean field Fokker-Planck equations including quantum effects.
\end{abstract}

\maketitle


\section{Introduction}

This paper continues our theoretical investigations (see \cite{grossmann} for a short review) of the barotropic Smoluchowski-Poisson (BSP) system (\ref{sp1})-(\ref{sp2}). For an isothermal equation of state $p=\rho k_B T/m$, these equations describe the mean field dynamics of a dissipative gas of self-gravitating Brownian particles in the overdamped limit $\xi\rightarrow +\infty$ where inertial effects are neglected. For the sake of generality, we shall consider an arbitrary barotropic equation of state  $p=p(\rho)$ so as to take  into account anomalous diffusion or short-range interactions. This type of equations appears in different contexts (with different interpretations) such as planetary formation in the solar nebula \cite{aaplanetes}, chemotaxis of biological populations \cite{ks}, colloids at a fluid interface driven by attractive capillary interactions \cite{dietrich} and nanoscience \cite{lebowitz,holm}. They also provide a simplified model of gravitational dynamics which displays many features common with ordinary self-gravitating systems such as collapse and evaporation \cite{grossmann}.  An interest of this model is that many analytical results can be derived since the inertia of the particles is neglected. For these reasons, the theoretical study of the BSP system is of considerable importance.

When studying equations of that type, it is natural to investigate
first the dynamical stability of a spatially homogeneous
distribution. However, in the gravitational case, we are rapidly
confronted to the well-known problem that a spatially homogeneous
distribution is not a steady state of these equations. To circumvent
this difficulty, one possibility is to advocate the ``Jeans swindle''
as in similar problems of astrophysics based on the
barotropic Euler-Poisson system \cite{jeans1,jeans2,bt}. This
procedure allows one to perform the stability analysis in a simple
manner. Although this first-step study certainly sheds light on the
problem, it is however not satisfactory and must be improved. Another
possibility is to study the dynamical stability of spatially
inhomogeneous steady states. This has been done in \cite{crs,sciso}
for isothermal spheres, in \cite{cspoly1,scpoly2} for polytropic
spheres and in \cite{virial1} for an arbitrary equation of state
by extending the methods developed in astrophysics
\cite{eddingtonstab,chandra,ebert,bonnoriso,mccreaiso,bonnorpoly,ledoux,antonov,lbw,yabushita,horedt,katz,paddystat,aaiso,aapoly,grand}.
Finally, a third possibility is to construct time-dependent solutions
of the BSP system.  Solutions describing the collapse, the
post-collapse and the evaporation of the system have been constructed
in
\cite{crs,sciso,post,virial1,cspoly1,scpoly2,childress,herrero,brenner,acedo,kavallaris,lushnikov}
by looking for self-similar, or close to self-similar, solutions. A
fully analytical solution of the BSP system has been obtained in
\cite{zero} in the cold case where the pressure can be
neglected. [Similar studies have been previously done
in astrophysics based on the barotropic Euler-Poisson system
\cite{hunter1,mestel,penston,larson, fillmore,bertschinger,lemos}. The results are,
however, different because the barotropic Euler-Poisson system takes
into account the inertia of the particles but neglects dissipation
while the BSP system is obtained in a strong friction limit in which
the inertia of the particles can be neglected. The  barotropic
Euler-Poisson system and the  BSP system correspond therefore to
opposite limits (weak and strong dissipation).]

Among these solutions, there is a very simple one, namely the collapse of a
homogeneous sphere. It is  found  \cite{zero} that the sphere undergoes a finite time singularity, resulting in
a vanishing radius and an infinite density in a finite time. The question that naturally emerges is
whether this time-dependent solution is stable with respect to small perturbations and, if not, how the perturbations grow. This  amounts to considering the development of density fluctuations
in a time-dependent  collapsing cloud. In a sense, this is the right formulation
of the Jeans problem without ``swindle''. We may note the analogy with the formation of stars in astrophysics through the collapse and the fragmentation of a molecular cloud \cite{hoyle,hunter1,hunter}. While the cloud collapses under its own gravity,
it also develops numbers of irregularities that grow and finally form stars. We may
also note the analogy with the formation of large-scale structures in cosmology \cite{bonnor} except that the universe is expanding \footnote{Of course, there also exists contracting solutions of the Friedmann  equations in cosmology corresponding to a closed universe \cite{peebles}.} while, in our problem, the cloud is contracting. In a sense, this corresponds to a universe experiencing a ``big-crunch''. To perform our study, we shall use many methods introduced in cosmology \cite{peebles,paddycosmo}. There will be, however,  two main differences: (i) our system is contracting instead of expanding; (ii) inertial effects are neglected in our model. The second point  is a huge simplification with respect to cosmology that allows us to solve the linear stability problem fully analytically. We can therefore easily  follow the growth of perturbations in the linear regime for an arbitrary equation of state. Furthermore, we derive an exact system of nonlinear equations [see Eqs. (\ref{c12})-(\ref{c13})] describing the formation of clusters on longer times. These equations are simpler than those considered in cosmology and could serve as a prototype to study the process of self-organization in complex media with structureless initial conditions.

The paper is organized as follows. In the first part of the paper (Secs. \ref{sec_spstat}-\ref{sec_ngc}), we consider the classical BSP system (\ref{sp1})-(\ref{sp2}). In Sec. \ref{sec_spstat}, we study the dynamical stability of an infinite homogeneous distribution in a static frame by invoking the ``Jeans swindle''. We find that the system is stable if and only if (iff) the wavelength of the perturbation  is smaller than the Jeans length $\lambda_J$. The perturbations decay exponentially rapidly if $\lambda<\lambda_J$ and  grow exponentially rapidly if $\lambda>\lambda_J$. In Sec. \ref{sec_coll}, we derive a new set of equations [see Eqs. (\ref{c12}), (\ref{c13}) and (\ref{h8})] describing the evolution of the density contrast in a homogeneous collapsing cloud.  We study the development of perturbations in the linear regime and obtain an analytical solution for the time evolution of the Fourier modes of the density contrast. For a polytropic equation of state, we find that the evolution depends on the value of the polytropic index $\gamma$. There exists a critical index $\gamma_{4/3}\equiv 2(d-1)/d$, already encountered  in our previous studies \cite{cspoly1,virial1,scpoly2} and well-known in astrophysics \cite{eddingtonstab,ledoux}, such that the system is stable for $\gamma>\gamma_{4/3}$ and unstable for  $\gamma<\gamma_{4/3}$. The effect of the Jeans length $\lambda_J(t)$, which is a function of time, manifests itself only in the initial stage and disappears as time goes on. In the unstable case, the perturbation is always growing when $\lambda>\lambda_J(0)$ while it starts decaying, before finally growing, when $\lambda<\lambda_J(0)$. We find that the perturbation grows algebraically for large times. This is in sharp contrast with the exponential growth predicted by the ``naive`'' Jeans analysis in a static medium. In the stable case, the perturbation is always decaying if $\lambda<\lambda_J(0)$ while it starts growing, before finally decaying, when $\lambda>\lambda_J(0)$.  For $\gamma=\gamma_{4/3}$, the Jeans length is constant. The perturbation grows for $\lambda>\lambda_J$ and decays for $\lambda<\lambda_J$. This is similar to the result of the Jeans analysis except that the evolution is algebraic instead of exponential. In Sec. \ref{sec_ngc}, we study the nonlinear regime and derive several analytical results. In particular, we determine the growth of the skewness and the long-wavelength tail of the power spectrum. We also find an exact self-similar solution to the nonlinear equations valid for large times. In the second part of the paper (Sec. \ref{sec_qsp}), we include quantum effects in our model. Specifically, we consider a self-gravitating Bose-Einstein condensate (BEC) in the presence of dissipative effects. Self-gravitating BECs have been proposed recently as a model of dark matter in cosmology \cite{bohmer}. Starting from a dissipative form of the Gross-Pitaevskii-Poisson system and using the Madelung transformation, we derive the quantum damped barotropic Euler-Poisson system. Considering a strong friction limit, we obtain the quantum barotropic Smoluchowski-Poisson system (\ref{mad16})-(\ref{mad17}). This corresponds to the classical  barotropic Smoluchowski-Poisson system  (\ref{sp1})-(\ref{sp2}) with an additional term called the Bohm quantum potential. We generalize our previous analysis in this more general context.

\section{Smoluchowski-Poisson system in a static frame}
\label{sec_spstat}

\subsection{Barotropic Smoluchowski-Poisson system}
\label{sec_sp}

We consider the barotropic Smoluchowski-Poisson (BSP) system
\begin{equation}
\label{sp1}
\xi\frac{\partial\rho}{\partial t}=\nabla\cdot   (\nabla p+\rho\nabla\Phi),
\end{equation}
\begin{equation}
\label{sp2}
\Delta\Phi=S_d G \rho,
\end{equation}
where $\rho({\bf r},t)$ is the spatial density of the particles, $\Phi({\bf r},t)$ is the gravitational potential, $p({\bf r},t)$ is the pressure, $\xi$ is the friction coefficient and $S_d$ is the surface of a unit sphere in $d$ dimensions. Basically, this equation describes the competition between pressure effects and gravity. To close the system, we assume that the pressure is related to the density by an arbitrary barotropic equation of state $p=p(\rho)$. The case of (ordinary) Brownian particles corresponds to an isothermal equation of state
\begin{equation}
\label{sp3}
p=\rho \frac{k_B T}{m},
\end{equation}
where the constant $T$ is the temperature of the bath and $m$ the mass of the particles ($k_B$ is the Boltzmann constant). Another very useful, and popular, equation of state is the polytropic equation of state
\begin{equation}
\label{sp4}
p=K\rho^{\gamma},\qquad \gamma=1+\frac{1}{n},
\end{equation}
where $K$ is the polytropic constant and $\gamma$, or $n$, is the polytropic index. The isothermal case is recovered for $\gamma=1$, i.e. $n\rightarrow \infty$, and $K=k_BT/m$. The velocity of sound $c_s$ is defined by the relation $c_s^2=p'(\rho)$. For an isothermal equation of state $c_s^2=k_BT/m$ is constant and for a polytropic equation of state $c_s^2=K\gamma\rho^{\gamma-1}$ depends on the density.

The Smoluchowski-Poisson system with an isothermal equation of state (\ref{sp3}), corresponding to self-gravitating Brownian particles, has been derived in \cite{hb2}. It is valid in a mean field approximation that becomes exact in a proper thermodynamic limit  $N\rightarrow +\infty$ such that the normalized temperature $\eta=\beta GMm/R^{d-2}$ is unity ($R$ denotes the size of the system and $\beta=1/k_B T$ the inverse temperature). The barotropic Smolchowski-Poisson system with an arbitrary equation of state $p(\rho)$ has been derived in \cite{gen,nfp} from a notion of generalized thermodynamics and in \cite{physicaA} from the dynamic density functional theory (DDFT). In the two approaches, the nonlinear pressure arises from short-range interactions between the particles but the physical arguments leading to Eq. (\ref{sp1}) are different.

\subsection{Jeans-type instability}
\label{sec_jeans}

A steady state of the BSP system (\ref{sp1})-(\ref{sp2}) satisfies the relation
\begin{equation}
\label{j1}
\nabla p+\rho\nabla\Phi={\bf 0},
\end{equation}
which can be viewed as a condition of hydrostatic balance. Since $p=p(\rho)$, this relation can be integrated to yield $\rho=\rho(\Phi)$. Using the Poisson equation (\ref{sp2}), we obtain the differential equation
\begin{equation}
\label{j2}
\Delta\Phi=S_d G \rho(\Phi),
\end{equation}
which has to be solved with appropriate boundary conditions. We note that self-gravitating Brownian particles described by the barotropic Smoluchowski-Poisson system have the same equilibrium states as barotropic stars described by the barotropic Euler-Poisson system \cite{bt,chandra}. However, their dynamics is different since the motion of Brownian particles is overdamped. The barotropic Smoluchowski equation (\ref{sp1}) can be written
\begin{equation}
\label{j2add1}
\xi\frac{\partial\rho}{\partial t}=\nabla\cdot \left (\rho\nabla\frac{\delta F}{\delta\rho}\right ),
\end{equation}
where $F$ is the free energy functional
\begin{equation}
\label{j2add2}
F=\frac{1}{2}\int\rho\Phi\, d{\bf r}+\int\rho\int^{\rho}\frac{p(\rho_1)}{\rho_1^2}\, d\rho_1\, d{\bf r}.
\end{equation}
The BSP system satisfies the $H$-theorem
\begin{equation}
\label{j2add3}
\dot F=\int \frac{\delta F}{\delta\rho}\frac{\partial\rho}{\partial t}\, d{\bf r}=-\frac{1}{\xi}\int\rho\left (\nabla\frac{\delta F}{\delta\rho}\right )^2\, d{\bf r}\le 0.
\end{equation}
Therefore, the BSP system relaxes towards the state that minimizes the free energy at fixed mass. This corresponds to the condition of hydrostatic equilibrium (\ref{j1}). In this sense, the BSP system is consistent with the statistical equilibrium state in the canonical ensemble. By contrast, the functional (\ref{j2add2}) is conserved by the barotropic Euler-Poisson system so that there is no {\it relaxation} towards equilibrium in that case \cite{bt}.

A natural problem concerns the dynamical stability of a steady state of the BSP system.   We shall restrict ourselves to linear dynamical stability. To that purpose, we consider a small perturbation $\delta\rho({\bf r},t)$ around a steady state $\rho({\bf r})$ of Eqs. (\ref{sp1})-(\ref{sp2}) and study its dynamical evolution. In the general case, the linearized BSP system is
\begin{equation}
\label{j3}
\xi\frac{\partial\delta\rho}{\partial t}=\nabla\cdot   \lbrack\nabla (c_s^2(\rho)\delta\rho)+\delta\rho\nabla\Phi+\rho\nabla\delta\Phi\rbrack,
\end{equation}
\begin{equation}
\label{j4}
\Delta\delta\Phi=S_d G \delta\rho.
\end{equation}
The dynamical stability of box-confined isothermal and polytropic configurations \footnote{When $d>2$, isothermal distributions and polytropic distributions  with $\gamma<\gamma_{6/5}\equiv 2d/(d+2)$ must be confined within a ``box'' otherwise they have infinite mass. In the context of chemotaxis, the box has a clear physical meaning since it represents the container in which the biological entities live.} has been studied in \cite{crs,sciso,cspoly1,scpoly2}. For example, for the isothermal equation of state in $d=3$, it is found that the system is stable if $\eta\equiv \beta GMm/R<\eta_c\simeq 2.52$  and if the density contrast ${\cal R}\equiv \rho(0)/\rho(R)<{\cal R}_c\simeq 32.1$; it is unstable if one of these two conditions is not met \cite{aaiso}. Similarly, for a polytropic equation of state in $d=3$, it is found that incomplete polytropes with $\gamma<6/5$ ($n>5$) are stable iff their inverse polytropic temperature $\eta\equiv M\lbrack 4\pi G/K(n+1)\rbrack^{n/(n-1)}/4\pi R^{(n-3)/(n-1)}$ and their density contrast ${\cal R}\equiv \rho(0)/\rho(R)$ are sufficiently small (the exact values of $\eta_c(n)$ and ${\cal R}_c(n)$ depend on the index $n$) \cite{aapoly}. The dynamical  stability of a general equation of state $p(\rho)$ for self-confined configurations  has been studied in \cite{virial1}  by adapting the methods of astrophysics \cite{eddingtonstab,ledoux}. For a polytropic equation of state in $d$ dimensions, it is found that a spatially inhomogeneous self-confined configuration is stable iff $\gamma\ge\gamma_{4/3}\equiv 2(d-1)/d$. When the steady state is unstable, it usually collapses or evaporates depending whether diffusion (pressure) or attraction (gravity) prevails \cite{grossmann}.

Let us now consider the case of a spatially homogeneous distribution. In that case, Eqs. (\ref{j3})-(\ref{j4}) reduce to
\begin{equation}
\label{j5}
\xi\frac{\partial\delta\rho}{\partial t}=c_s^2\Delta\delta\rho+S_d G\rho\delta\rho,
\end{equation}
where $c_s^2=c_s^2(\rho)$ is constant. Considering for simplicity an infinite system \footnote{This is not necessary and Eq. (\ref{j5}) could be solved in a box by decomposing the perturbations on the eigenmodes of the Laplacian \cite{cd}.} and decomposing the perturbations in plane waves of the form $e^{i({\bf k}\cdot {\bf r}-\omega t)}$, we obtain the dispersion relation
\begin{equation}
\label{j6}
i\xi\omega=c_s^2k^2-S_dG\rho.
\end{equation}
This relation, like the original Smoluchowski equation (\ref{sp1}), clearly shows the competition between the attractive gravitational force and the repulsive pressure (assuming $c_s^2>0$). There exists a critical wavenumber
\begin{equation}
\label{j7}
k_J=\left (\frac{S_dG\rho}{c_s^2}\right )^{1/2},
\end{equation}
corresponding to the Jeans wavenumber in astrophysics \cite{jeans1,jeans2,bt}. However, contrary to the astrophysical problem based on the barotropic Euler-Poisson system, the evolution of the perturbations is different in our case. This is because the Smoluchowski-Poisson system is an overdamped model where inertial effects are neglected. Here, the perturbations evolve like $e^{\gamma t}$ with an exponential rate $\gamma=(S_dG\rho-c_s^2k^2)/\xi$. For $k<k_J$, the system is unstable and the perturbations grow with a growth rate $\gamma>0$; for $k>k_J$ the system is stable and the perturbations decay with a damping rate $\gamma<0$.  Note that the growth rate is maximum for $k=0$ (corresponding to an infinite wavelength $\lambda=2\pi/k\rightarrow +\infty$) and its value is $\gamma_{max}=S_dG\rho/\xi$. {\it In conclusion, Jeans' stability criterion is $k>k_J$ (i.e. $\lambda<\lambda_J$).}

Although this stability analysis, like the original Jeans study in astrophysics \cite{jeans1,jeans2}, is simple and sheds light on the problem, it however encounters a serious problem. Indeed, if $\rho$ is uniform (in a finite or infinite domain), the pressure gradient in Eq. (\ref{j1}) vanishes implying that the gravitational field must also vanish ($\nabla\Phi={\bf 0}$). Then, the Poisson equation (\ref{j2}) cannot be satisfied for $\rho>0$. This implies that there is no steady state of Eqs. (\ref{j1})-(\ref{j2}) with uniform density. Therefore, when using the perturbation equation (\ref{j5}), we make what is traditionally called the ``Jeans swindle'' \cite{bt}. There are different ways to circumvent this difficulty. (i) The first is to restrict ourselves to spatially inhomogeneous steady states that are solution of Eqs. (\ref{j1})-(\ref{j2}). (ii) The second is to consider a uniformly rotating system. In that case, if we introduce an effective gravitational potential $\Phi_{eff}=\Phi-({\bf \Omega}\times {\bf r})^2/2$ in the rotating frame, the Poisson equation is replaced by $\Delta\Phi_{eff}=4\pi G\rho-2\Omega^2$ and there exists spatially homogeneous distributions with $\rho=\Omega^2/2\pi G$. (iii) The third possibility is to modify the Poisson equation. This is justified for some physical systems. For example, in the context of chemotaxis \cite{ks}, the Poisson equation is replaced by a generalized field equation of the form $\chi\partial_t\Phi=\Delta\Phi-k_R^2\Phi-S_d G\rho$ where $k_R^{-1}$ is a screening length. This leads to the general Keller-Segel model. In that case, a spatially homogeneous distribution is a steady state of the equations with $\Phi=-S_d G\rho/k_R^2$. In some approximations, the generalized field equation $\chi\partial_t\Phi=\Delta\Phi-k_R^2\Phi-S_d G\rho$ is replaced by the screened Poisson equation $\Delta\Phi-k_R^2\Phi=S_d G\rho$ or by the modified Poisson equation $\Delta\Phi=S_d G(\rho-\overline{\rho})$ where $\overline{\rho}$ is the average density. Again, a spatially homogeneous distribution  satisfying $\Phi=-S_d G\rho/k_R^2$ or $\rho=\overline{\rho}$ respectively is a steady state of these equations. Therefore, there is no ``Jeans swindle'' in the chemotactic problem \cite{cjc,chemojeans,cd}. In gravity models, we could introduce a cosmological constant $\Lambda$ and replace the Poisson equation by $\Delta\Phi=S_dG\rho-\Lambda$. In that case, the particular homogeneous density distribution $\rho=\Lambda/S_d G$ is a steady state of the equations \footnote{As is well-known,  Einstein \cite{einstein} introduced a cosmological constant in the equations of general relativity in order to obtain a static homogeneous and isotropic universe.  As a preamble of his paper, he considered a modification of the classical Poisson equation in the form $\Delta\Phi-k_R^2\Phi=4\pi G\rho$. He incorrectly believed that, in the Newtonian world model, the cosmological constant is equivalent to a screening length. The fact that the cosmological constant leads to a Poisson equation of the form $\Delta\Phi=4\pi G\rho-\Lambda$ was understood later by Lema\^itre and Eddington  (see \cite{spiegel,kiessling,cd} for historical details). On the other hand, it was realized that the Einstein universe is strongly unstable \cite{eddington,harrison} so that the idea of a dynamical (expanding) universe finally emerged.}. However, this justification does not apply to a uniform mass of gas of different density. Furthermore, this homogeneous distribution is unstable as shown in Appendix \ref{sec_cc}. (iv) Kiessling \cite{kiessling} has provided a vindication of the Jeans swindle. He argues that, when considering an infinite and homogeneous distribution of matter, the Poisson equation must be modified so as to correctly define the gravitational force. He proposes to use a regularization of the form $\Delta\Phi-k_R^2\Phi=4\pi G\rho$ where $k_R$ is an inverse screening length that ultimately tends to zero ($k_R\rightarrow 0$), or a regularization of the form  $\Delta\Phi=4\pi G(\rho-\overline{\rho})$ where $\overline{\rho}$ is the mean density. In his point of view, the modification of the Poisson equation is not a swindle but just the right way to make the problem mathematically rigorous and correctly define the gravitational force. However, this argument applies only to an infinite system. For a finite system, a spatially homogeneous distribution is not a steady state of the BSP equations.

{\it Remark:} a form of Jeans criterion can be recovered for box-confined configurations. For example, for finite isothermal spheres in $d=3$, introducing the mean density $\overline{\rho}=3M/4\pi R^3$,  the stability criterion $\eta=\beta GMm/R\le \eta_c\simeq 2.52$  \cite{aaiso} can be rewritten  $R<R_J\equiv  (3\eta_c/4\pi G\overline{\rho}m \beta)^{1/2}$ where $R_J$ is similar to the Jeans length $\lambda_J=2\pi (k_BT/4\pi G\rho m)^{1/2}$ corresponding to Eq. (\ref{j7}). Similarly, for finite polytropic spheres  with $n>5$, the stability criterion $\eta\equiv M\lbrack 4\pi G/K(n+1)\rbrack^{n/(n-1)}/4\pi R^{(n-3)/(n-1)}<\eta_c(n)$ \cite{aapoly} can be rewritten $R<R_J\equiv \lbrack K(n+1)\phi(n)/4\pi Gn\rho^{(n-1)/n}\rbrack^{1/2}$ with $\phi(n)=n\lbrack 3\eta_c(n)\rbrack^{(n-1)/n}$ where $R_J$ is similar to the Jeans length $\lambda_J=2\pi \lbrack K(n+1)/4\pi Gn\rho^{(n-1)/n}\rbrack^{1/2}$ corresponding to Eq. (\ref{j7}).

\subsection{Homogeneous collapsing cloud}
\label{sec_hom}

It is clear that a finite spatially homogeneous solution of the BSP system will collapse under its own gravity. Similarly, a finite spatially homogeneous solution of the Euler-Poisson system is not steady and will evolve in time. However, since the Euler-Poisson system is an inertial model, depending on its energy, the system will either expand (positive energy) or collapse (negative energy). This is at the basis of the Newtonian cosmological models of McCrea \& Milne \cite{mcCreamilne} where the universe is viewed as a homogeneous self-gravitating sphere of radius $a(t)$ \footnote{It turns out that the equations of Newtonian cosmology are identical with the Friedmann equations derived from the theory of general relativity, provided the pressure is negligible in comparison with the energy density $\rho c^2$ where $c$ is the speed of light. It is surprising to realize that Newtonian cosmology was developed after, and was influenced by, the theory of general relativity while it could have been introduced much earlier.}. In cosmology, we can have open (always expanding) or closed (expanding then contracting) models of the universe with an intermediate case provided by the Einstein-de Sitter (EdS) universe \cite{weinberg,peebles}. In our problem, since the velocity of the particles is directly proportional to the gravitational force (overdamped limit), a homogeneous cloud can only collapse under its own gravity. In the analogy with cosmology, this corresponds to a ``big-crunch''. This is also  similar to the gravitational collapse of a molecular cloud in relation to the process of stars formation \cite{hoyle,hunter1,hunter}.

It is easy to construct a time-dependent spatially homogeneous solution of the BSP system. For a spatially homogeneous distribution of the form
\begin{equation}
\label{h1}
\rho({\bf r},t)=\rho_b(t),
\end{equation}
the BSP system (\ref{sp1})-(\ref{sp2}) reduces to
\begin{equation}
\label{h2}
\frac{d\rho_b}{dt}=\frac{S_d G}{\xi}\rho_b^2.
\end{equation}
The pressure $p$ does not appear in the theory of the homogeneous model since it enters Eq. (\ref{sp1}) only through its gradient. Therefore, we recover the equation of a cold system ($p=0$) considered in \cite{zero}. If we consider the spherically symmetric collapse of a uniform sphere of radius $a(t)$, the conservation of mass imposes
\begin{equation}
\label{h3}
M=\frac{1}{d}S_d\rho_b a^d.
\end{equation}
Solving Eq. (\ref{h2}) and using Eq. (\ref{h3}), we find that the density and the radius of a collapsing  homogeneous sphere evolve according to
\begin{equation}
\label{h4}
\rho_b(t)=\frac{\rho_b(0)}{1-\frac{t}{t_*}},
\end{equation}
\begin{equation}
\label{h5}
a(t)=a(0)\left (1-\frac{t}{t_*}\right )^{1/d},
\end{equation}
where
\begin{equation}
\label{h6}
t_*=\frac{\xi}{S_d G\rho_b(0)}=\frac{\xi a(0)^d}{dGM}.
\end{equation}
This solution leads to a finite time collapse. At $t=t_*$, all the mass is in a Dirac peak at $r=0$ (big crunch). This solution is the counterpart of the homogeneous solution found by Hunter \cite{hunter1}, Mestel \cite{mestel} and Penston \cite{penston}  in the inertial case.

We can also obtain these equations from the Lagrangian motion of a fluid particle as in \cite{zero} by using the fact that the Smoluchowski  equation (\ref{sp1}) with $\nabla p={\bf 0}$ can be interpreted as a continuity equation with a velocity field
\begin{equation}
\label{h7}
{\bf u}\equiv \frac{d{\bf r}}{dt}=-\frac{1}{\xi}\nabla\Phi.
\end{equation}
Applying this equation  at the border of the sphere, and using the Gauss theorem, we obtain
\begin{equation}
\label{h8}
\xi\frac{da}{dt}=-\frac{GM}{a^{d-1}}=-\frac{1}{d}S_dG\rho_b a,
\end{equation}
which after integration returns Eq. (\ref{h5}). In the analogy with
cosmology, Eqs. (\ref{h3}) and (\ref{h8}) are the counterparts of the
Friedmann equations when pressure effects are
neglected. We note, however, that the Friedmann
equations are second order in time while Eq. (\ref{h8}) is first
order.

\section{Smoluchowski-Poisson system in a collapsing frame}
\label{sec_coll}

\subsection{The fundamental equations}
\label{sec_fund}

We shall now study the instability of the collapsing homogeneous sphere and the development of irregularities. This process of fragmentation, driven by gravity, ultimately leads to the formation of localized dense clusters similar to galaxies in the universe or stars in a molecular cloud. Like in cosmology, we shall work in the comoving frame (here the collapsing frame). To that purpose, we set
\begin{equation}
\label{c1}
{\bf r}=a(t){\bf x},
\end{equation}
where $a(t)$ is the scale factor (in our problem it corresponds to the radius of the sphere). Let us first write the Poisson equation (\ref{sp2}) in the collapsing frame. For the background distribution (\ref{h1}), the Gauss theorem reads
\begin{equation}
\label{c2}
\frac{d\Phi_b}{dr}=\frac{GM(r)}{r^{d-1}}=\frac{1}{d}S_dG\rho_b r=-\xi\frac{\dot a}{a}r,
\end{equation}
where $M(r)$ is the mass enclosed within the sphere of radius $r$ and we have used Eq. (\ref{h8}) to obtain the last equality. The background gravitational potential is therefore
\begin{equation}
\label{c3}
\Phi_b=-\frac{1}{2}\xi\frac{\dot a}{a}  r^2=-\frac{1}{2}\xi {\dot a}a x^2.
\end{equation}
In terms of the new potential $\phi=\Phi-\Phi_b$, i.e.
\begin{equation}
\label{c4}
\phi=\Phi+\frac{1}{2}\xi {\dot a}a x^2,
\end{equation}
the Poisson equation (\ref{sp2}) becomes
\begin{equation}
\label{c5}
\Delta\phi=S_d G a^2 (\rho-\rho_b).
\end{equation}
In this expression, the Laplacian is defined with respect to the variable ${\bf x}$  and we have used Eq. (\ref{h8}) to obtain the second term on the r.h.s.

We now write the barotropic Smoluchowski equation  in the collapsing frame. We first note that
\begin{equation}
\label{c6}
\left (\frac{\partial}{\partial t}\right )_r\rho({\bf r}/a(t),t)=\frac{\partial\rho}{\partial t}-\frac{\dot a}{a}{\bf x}\cdot\nabla\rho.
\end{equation}
Therefore, the barotropic Smoluchowski equation (\ref{sp1}) becomes
\begin{equation}
\label{c7}
\frac{\partial\rho}{\partial t}-\frac{\dot a}{a}{\bf x}\cdot\nabla\rho=\frac{1}{\xi a^2}\nabla\cdot  (\nabla p+\rho\nabla\Phi),
\end{equation}
where the derivatives are taken with respect to ${\bf x}$. Introducing the new potential (\ref{c4}),  we obtain after simplification
\begin{equation}
\label{c8}
\frac{\partial\rho}{\partial t}+d\frac{\dot a}{a}\rho=\frac{1}{\xi a^2}\nabla\cdot (\nabla p+\rho\nabla\phi).
\end{equation}
Like in cosmology, it is convenient to write the density in the form
\begin{eqnarray}
\label{c9}
\rho=\rho_b(t)\left\lbrack 1+\delta({\bf x},t)\right \rbrack,
\end{eqnarray}
where $\rho_b\propto 1/a^d$ according to Eq. (\ref{h3}) and $\delta({\bf x},t)$ is the density contrast \cite{peebles}. Substituting Eq. (\ref{c9})  in Eq. (\ref{c8}), we obtain the barotropic Smoluchowski equation in the collapsing frame
\begin{equation}
\label{c10}
\rho_b\frac{\partial\delta}{\partial t}=\frac{1}{\xi a^2}\nabla\cdot ( \nabla p+\rho_b(1+\delta)\nabla\phi).
\end{equation}
Since $p=p(\rho)$, this can be written
\begin{equation}
\label{c11}
\frac{\partial\delta}{\partial t}=\frac{1}{\xi a^2}\nabla\cdot ( c_s^2\nabla \delta+(1+\delta)\nabla\phi),
\end{equation}
where $c_s^2=p'(\rho)=p'(\rho_b (1+\delta))$ is the square of the velocity of sound in the evolving system. It generically depends on position and time (it is constant only for an isothermal equation of state).  Finally, substituting Eq. (\ref{c9}) in the Poisson equation (\ref{c5}), we obtain our final system of equations
\begin{equation}
\label{c12}
\xi\frac{\partial\delta}{\partial t}=\frac{1}{a^2}\nabla\cdot (c_s^2\nabla \delta+(1+\delta)\nabla\phi),
\end{equation}
\begin{equation}
\label{c13}
\Delta\phi=S_d G \rho_b \delta a^2,
\end{equation}
where $\rho_b(t)$ and $a(t)$ are given by Eqs. (\ref{h4}) and (\ref{h5}). If we measure the evolution in terms of $a$ instead of $t$, we can rewrite Eq. (\ref{c12}) in the form
\begin{equation}
\label{c14}
\frac{\partial\delta}{\partial a}=-\frac{d}{S_dG\rho_b a^3}\nabla\cdot (c_s^2\nabla \delta+(1+\delta)\nabla\phi),
\end{equation}
where we have used Eq. (\ref{h8}). For an isothermal equation of state, Eq. (\ref{c12}) becomes
\begin{equation}
\label{c15}
\xi\frac{\partial\delta}{\partial t}=\frac{1}{a^2} \left (\frac{k_B T}{m}\Delta\delta+S_d G \rho_b \delta (1+\delta) a^2+\nabla\delta\cdot \nabla\phi\right ),
\end{equation}
where we have  used  the Poisson equation (\ref{c13}) to get the second term on the r.h.s. For a cold system $c_s=0$, we get
\begin{equation}
\label{c16}
\xi\frac{\partial\delta}{\partial t}= S_d G \rho_b \delta (1+\delta)+\frac{1}{a^2}\nabla\delta\cdot \nabla\phi.
\end{equation}
Equations (\ref{c12})-(\ref{c13}) could form a prototypical model to study the formation and the growth of structures in complex media with structureless initial condition. As we shall see, they lead to a process of fragmentation and  self-organization. Furthermore, they are simpler to study than the equations used in cosmology \cite{peebles} since there is no inertia in our model. Indeed, Eqs. (\ref{c12})-(\ref{c13}) are parabolic while the equations used in cosmology are hyperbolic. We therefore expect to obtain simpler, hopefully analytical, results like in the study of the BSP system (\ref{sp1})-(\ref{sp2}).

\subsection{Linearized equations}
\label{sec_lin}

Of course, $\delta=\phi=0$ is a particular solution of Eqs. (\ref{c12})-(\ref{c13}) corresponding to the pure background flow (\ref{h1}). We shall now study the dynamical stability of this solution. If we consider small perturbations $\delta\ll 1$, $\phi\ll 1$ around that state, we obtain the linearized equation
\begin{equation}
\label{lin1}
\xi\frac{\partial\delta}{\partial t}=\frac{c_s^2}{a^2}\Delta\delta+S_d G \rho_b \delta,
\end{equation}
where $c_s^2=p'(\rho_b)$ is the square of the velocity of sound in the background flow. It depends only on time. Equation (\ref{lin1}) is the counterpart of the equation derived by Bonnor \cite{bonnor} in cosmology. Expanding the solutions of Eq. (\ref{lin1}) in Fourier modes of the form $\delta({\bf x},t)=\delta_{\bf k}(t)e^{i{\bf k}\cdot {\bf x}}$, we find that
\begin{eqnarray}
\label{lin2}
\xi\dot\delta+\left (\frac{c_s^2k^2}{a^2}-S_d G\rho_b\right )\delta=0,
\end{eqnarray}
where, for convenience, we have noted $\delta$ for $\delta_{\bf k}$. From this equation, we can define a time-dependent Jeans wavenumber in the comoving frame
\begin{eqnarray}
\label{lin3}
k_J=\left (\frac{S_d G\rho_b a^2}{c_s^2}\right )^{1/2}.
\end{eqnarray}
Note that the proper wavenumber obtained by writing $\delta\propto e^{i{\bf k}_*\cdot {\bf r}}$ is $k_J^*=\left ({S_d G\rho_b}/{c_s^2}\right )^{1/2}$. The evolution of $k_J(t)$ depends on the equation of state. For an isothermal equation of state for which $c_s^2=k_BT/m$, we can write  $k_J=\kappa_J a^{-(d-2)/2}$ where $\kappa_J=(S_d G\rho_b a^d m/k_BT)^{1/2}$ is a constant according to Eq. (\ref{h3}). The Jeans length $\lambda_J=2\pi/k_J$ behaves like $a^{(d-2)/2}$. For $d>2$ (resp. $d<2$), it decreases (resp. increases) with time so that the system becomes unstable at smaller and smaller (resp. larger and larger) scales as the cloud collapses. For $d=2$, the Jeans length is constant: $k_J=\kappa_J$. For a polytropic equation of state for which $c_s^2=K\gamma \rho_b^{\gamma-1}$, we can write $k_J=\kappa_J a^{(d\gamma-2(d-1))/2}$ where $\kappa_J=\lbrack S_d G(\rho_b a^d)^{2-\gamma}/K\gamma\rbrack^{1/2}$ is constant according to Eq. (\ref{h3}). The Jeans length behaves like $\lambda_J\propto a^{(2(d-1)-d\gamma)/2}$. For $\gamma<\gamma_{4/3}\equiv 2(d-1)/d$, the Jeans length decreases with time and, for $\gamma>\gamma_{4/3}$, it increases with time. For $\gamma=\gamma_{4/3}$, the Jeans length is constant: $k_J=\kappa_J$. The evolution of the proper Jeans length $\lambda_J^*$ is different but the evolution of the comoving Jeans length $\lambda_J$  is better suited to our problem.

{\it Remark:} our analysis exhibits a critical index $\gamma_{4/3}\equiv 2(d-1)/d$, the same as the one determining the dynamical stability of spatially inhomogeneous polytropic spheres \cite{cspoly1,virial1,scpoly2} (see also \cite{eddingtonstab,ledoux} in astrophysics).  It is interesting to note that this critical index does not appear when the Jeans problem is (incorrectly) formulated in a static homogeneous background (see Sec. \ref{sec_jeans}) \footnote{The index $\gamma_{4/3}\equiv 2(d-1)/d$ only appears when we introduce the Jeans mass $M_J\propto\rho \lambda_J^d\propto \rho^{(d\gamma-2(d-1))/2}$, but it does not play any particular role in the stability analysis of the homogeneous fluid.} while it appears when the Jeans problem is properly formulated in a collapsing background. This will have important consequences in the stability problem (see below).

\subsection{Solution of the equation for the density contrast}
\label{sec_sol}

The evolution of the perturbations is more complicated to analyze in a collapsing homogeneous medium than in a static homogeneous medium (and turns out to be very different). An interest of the present model is that the equation (\ref{lin2}) for the density contrast in the linear regime  can be solved analytically for any equation of state while the corresponding equation in cosmology can be solved analytically only in particular cases (this is because it is a second order differential equation in time while Eq. (\ref{lin2}) is a first order differential equation) \cite{peebles}.  Measuring the evolution in terms of $a$ instead of $t$ and using Eq. (\ref{h8}), we can rewrite Eq. (\ref{lin2}) in the form
\begin{eqnarray}
\label{sol1}
\frac{d\delta}{da}-\frac{d}{a}\left (\frac{c_s^2k^2}{S_d G\rho_b a^2}-1\right )\delta=0.
\end{eqnarray}
Its general solution is
\begin{eqnarray}
\label{sol1b}
\delta(a)\propto \frac{1}{a^{d}} e^{\frac{dk^2}{S_d G}\int^a\frac{c_s^2(a')}{\rho_b(a'){a'}^2}\, da'}.
\end{eqnarray}
For a cold gas $c_s=0$ in which the pressure is zero, Eq. (\ref{sol1}) reduces to
\begin{eqnarray}
\label{sol2}
\frac{d\delta}{da}+\frac{d}{a}\delta=0,
\end{eqnarray}
and its solution is
\begin{eqnarray}
\label{sol3}
\delta\propto a^{-d}\propto (1-t/t_*)^{-1}.
\end{eqnarray}
In that case, the homogeneous cloud is unstable at all scales and we can write in full generality $\delta({\bf x},a)=D({\bf x})/a^d$ or  $\delta({\bf x},t)=\delta({\bf x},0)(1-t/t_*)^{-1}$. We note that the growth of the perturbation is algebraic while the static Jeans study predicts an exponential growth with a rate $\gamma=S_d G\rho/\xi$ (see Sec. \ref{sec_jeans}). This is a very important difference. Assuming now a polytropic equation of state and introducing the notations of Sec. \ref{sec_lin}, we can rewrite Eq. (\ref{sol1}) in the form
\begin{eqnarray}
\label{sol4}
\frac{d\delta}{da}-\frac{d}{a}\left (\frac{k^2}{\kappa_J^2a^{d\gamma-2(d-1)}}-1\right )\delta=0.
\end{eqnarray}

\begin{figure}[!h]
\begin{center}
\includegraphics[clip,scale=0.3]{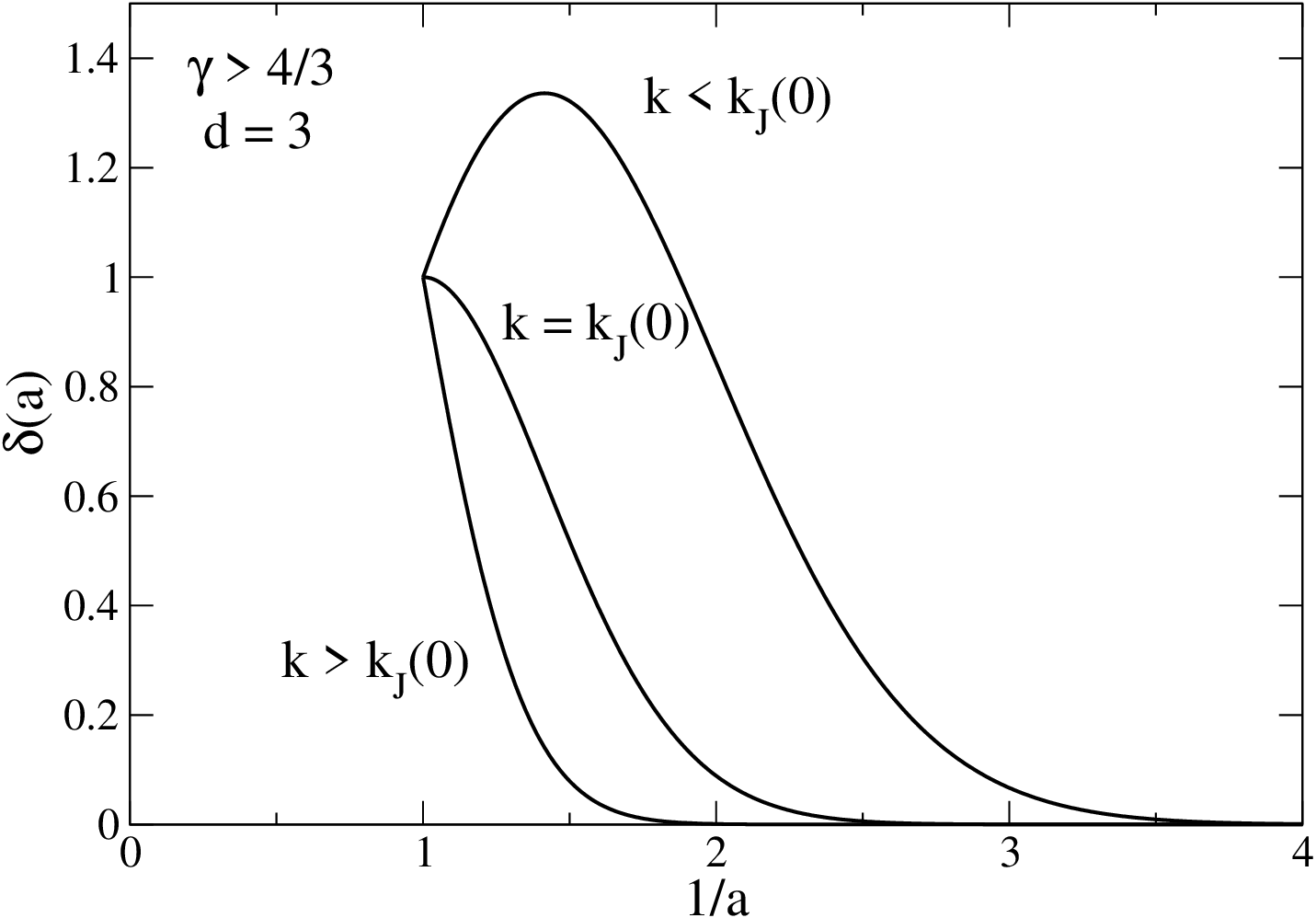}
\caption{Evolution of the perturbation $\delta(a)$ for $\gamma>4/3$ ($d=3$, $a_0=1$ and  $\delta_0=1$). We have taken $\gamma=2$ which corresponds to a BEC with quartic self-interactions in the TF approximation (see Sec. \ref{sec_eos}). The system is asymptotically stable. For $k<k_J(0)$, the perturbation grows before decaying. The maximum occurs at $a=(k/k_J)^{2/(3\gamma-4)}$ and its value $\delta_{max}\propto k^{-6/(3\gamma-4)}$ is maximum for small $k$ (large wavelengths).  }
\label{deltagammasup4sur3}
\end{center}
\end{figure}

\begin{figure}[!h]
\begin{center}
\includegraphics[clip,scale=0.3]{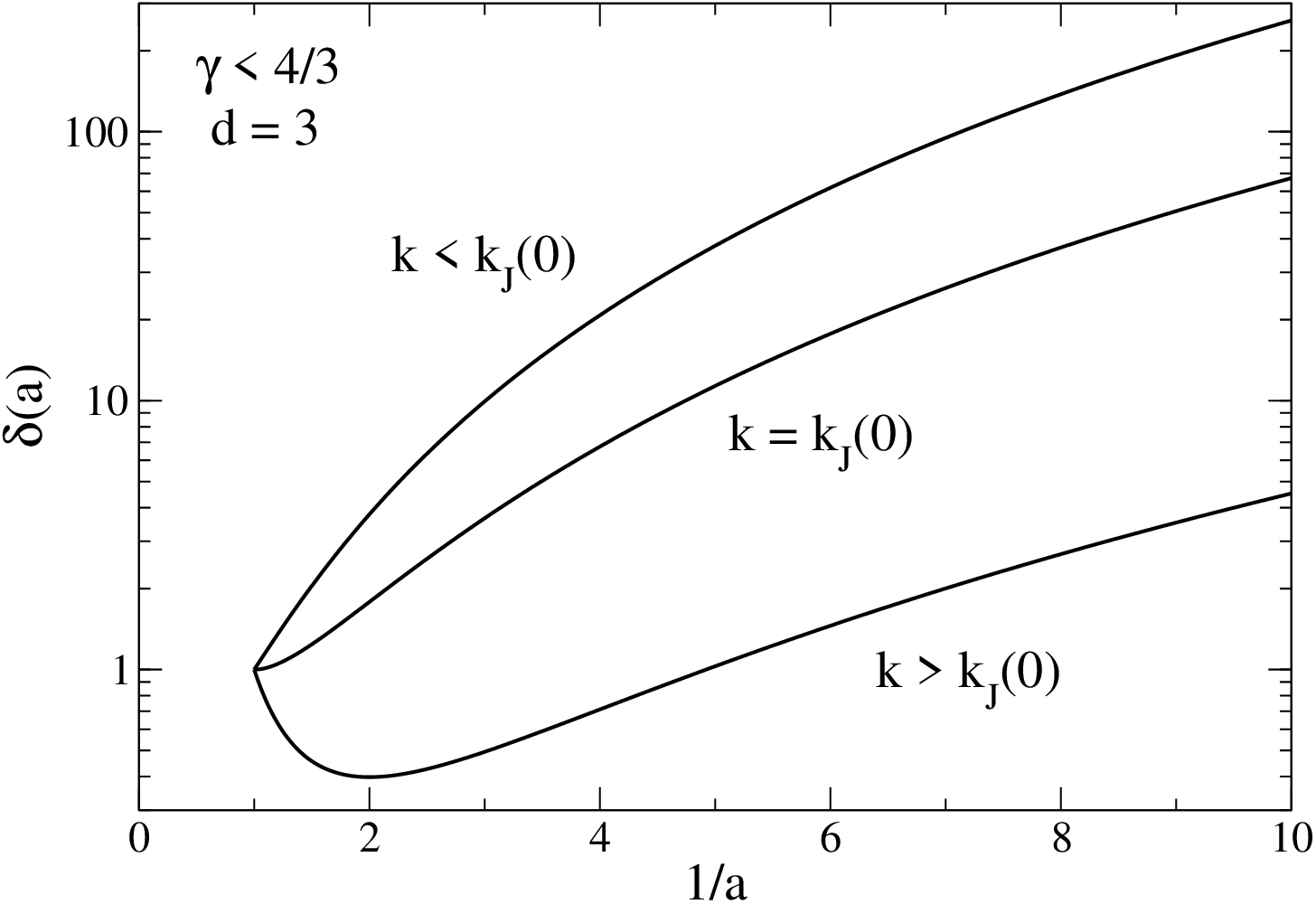}
\caption{Evolution of the perturbation $\delta(a)$ for $\gamma<4/3$ ($d=3$, $a_0=1$ and  $\delta_0=1$). We have taken $\gamma=1$ which corresponds to a classical self-gravitating Brownian gas (isothermal distribution). The system is asymptotically unstable. For $k>k_J(0)$, the perturbation decays before growing. The minimum occurs at $a=(k/k_J)^{-2/(4-3\gamma)}$ and its value $\delta_{min}\propto k^{6/(4-3\gamma)}$ is minimum for large $k$ (small wavelengths).}
\label{deltagammainf4sur3}
\end{center}
\end{figure}

\begin{figure}[!h]
\begin{center}
\includegraphics[clip,scale=0.3]{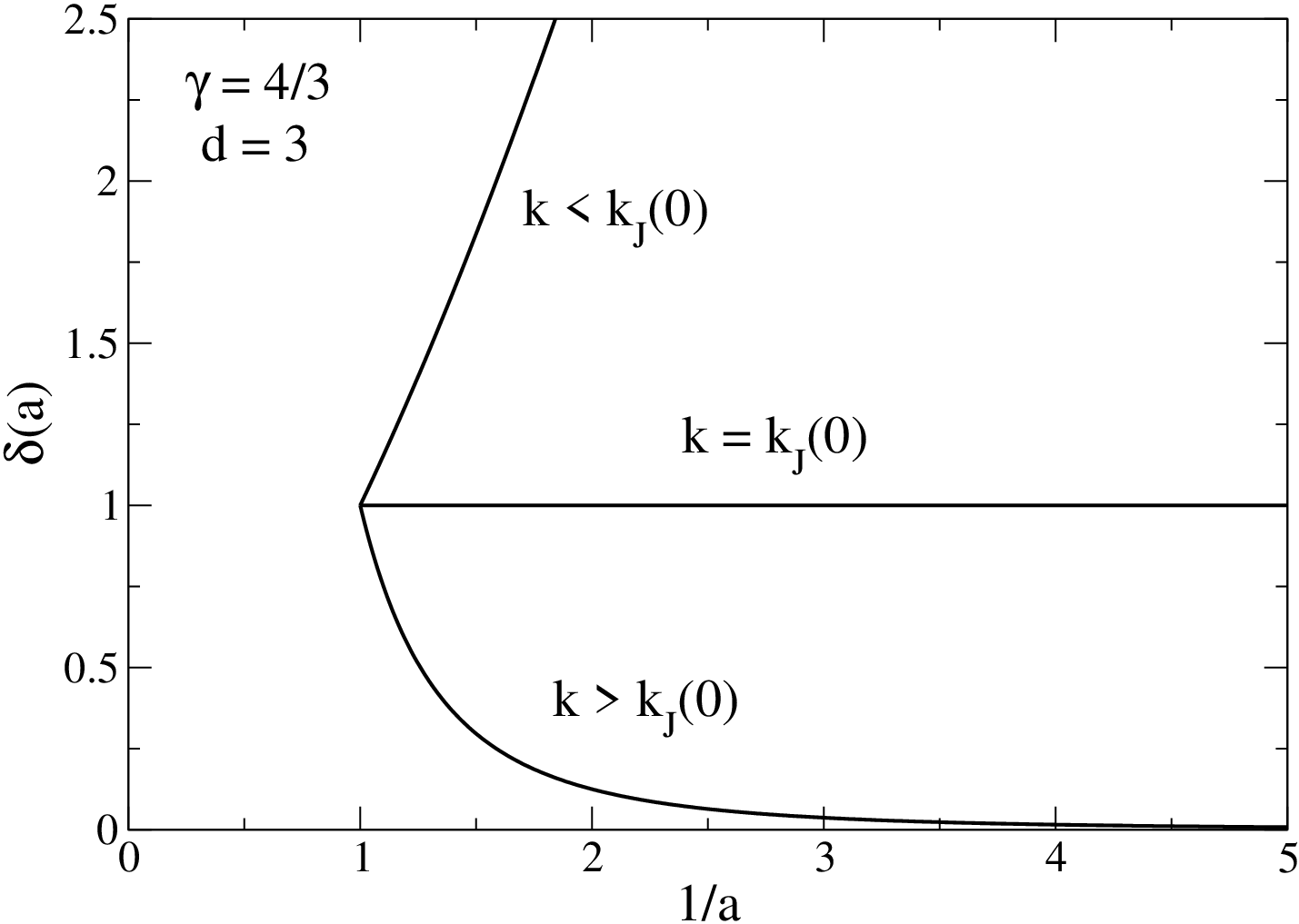}
\caption{Evolution of the perturbation $\delta(a)$ for $\gamma=4/3$. The system is stable for $k>k_J$ and unstable for $k<k_J$.  We have taken $a_0=\delta_0=1$.}
\label{deltagammaegal4sur3}
\end{center}
\end{figure}

If $\gamma\neq \gamma_{4/3}$ (i.e. $n\neq n_3$), the solution of this equation is
\begin{eqnarray}
\label{sol5}
\delta (a)\propto \frac{1}{a^d}e^{\frac{dk^2a^{2(d-1)-d\gamma}}{\kappa_J^2 (2(d-1)-d\gamma)}}.
\end{eqnarray}
This includes the isothermal equation of state ($\gamma=1$) with $d\neq 2$ as a particular case
\begin{eqnarray}
\label{sol6}
\delta (a)\propto \frac{1}{a^d}e^{\frac{dk^2a^{d-2}}{(d-2)\kappa_J^2}}.
\end{eqnarray}
For short times, writing $a=a_0(1-\epsilon)$ with $\epsilon\ll 1$, we obtain
\begin{eqnarray}
\label{sol7}
\frac{\delta (a)}{\delta (a_0)}\simeq 1+d\left (1-\frac{k^2}{k_J(0)^2}\right )\epsilon.
\end{eqnarray}
This equation clearly shows the initial effect of the Jeans wavelength. Indeed, the perturbation starts to grow when $k<k_J(0)$ and starts to decay when $k>k_J(0)$. This result, which is independent on the value of $\gamma$, is similar to the one obtained from the static Jeans study (see Sec. \ref{sec_jeans}). Indeed, for short times, the system does not ``see'' that the background is contracting. By contrast, for large times, the effect of the Jeans scale disappears and the evolution of the system is only controlled by the value of $\gamma$. For $\gamma>\gamma_{4/3}$, $\delta(a)\rightarrow 0$ very rapidly as $a\rightarrow 0$. The system becomes asymptotically stable to all wavelengths. For $\gamma<\gamma_{4/3}$,  $\delta(a)\propto a^{-d}\rightarrow +\infty$ as $a\rightarrow 0$. The system becomes asymptotically unstable to all wavelengths and behaves like in a cold gas. {\it Therefore, the system is asymptotically stable for $\gamma>\gamma_{4/3}$ and asymptotically unstable for $\gamma<\gamma_{4/3}$.} This stability criterion is very different from the Jeans stability criterion in a static background (see Sec. \ref{sec_jeans}) where the value of the index $\gamma$ does not play any crucial role. It is, on the other hand, compatible with the stability criterion of self-confined inhomogeneous polytropic spheres \cite{virial1}.

If $\gamma=\gamma_{4/3}$ (i.e. $n=n_3$), the solution of Eq. (\ref{sol4}) is
\begin{eqnarray}
\label{sol8}
\delta (a)\propto a^{d\left (k^2/\kappa_J^2-1\right )}.
\end{eqnarray}
This includes the isothermal equation of state ($\gamma=1$) with $d=2$ as a particular case. We see that the Jeans scale $k_J=\kappa_J$ now plays a crucial role at all times.  For $k<\kappa_J$, the perturbation increases in time and the system is unstable (for $k\ll \kappa_J$, the perturbation grows like in a cold gas). For $k>\kappa_J$, the perturbation decreases in time and the system is stable. These results are similar to those obtained in the static Jeans study except that the growth of the perturbation is algebraic instead of being exponential. {\it For the critical index $\gamma=\gamma_{4/3}$, the system is stable for $k>k_J$ and unstable for $k<k_J$.}

Some curves representing the evolution of the density contrast $\delta(a)$ in the different cases described above are represented in Figs. \ref{deltagammasup4sur3}-\ref{deltagammaegal4sur3}   for illustration.

{\it Remark 1:} for $\gamma>\gamma_{4/3}$, the system is asymptotically stable. However, for $k<k_J(0)$, the perturbations start  growing  before decaying. It can happen that their growth  generates nonlinear effects that may trigger new instabilities. Therefore, their full stability is not granted. It is likely that the nonlinear evolution leads to a process of fragmentation marked by the formation of dense localized clusters in {\it hydrostatic equilibrium} (virialized structures). These structures correspond to complete polytropic spheres that are dynamically stable since $\gamma>\gamma_{4/3}$ (see Sec. \ref{sec_jeans}).

{\it Remark 2:} it can be relevant to also treat the case where $c_s^2<0$ (corresponding to $\kappa_J^2<0$). In certain cases, an imaginary velocity of sound  arises from a negative pressure due to attractive short-range interactions (see Sec. \ref{sec_eos}). In that case, we find that the perturbations  $\delta(a)$ always grow. Therefore, a negative value of $c_s^2$ can enhance the gravitational instability and make the system very unstable.

\section{Nonlinear gravitational clustering}
\label{sec_ngc}

In this section, using an analogy with cosmology, we provide analytical results valid in the nonlinear regime.

\subsection{Second order perturbation theory for $\delta\rho/\rho_b$: skewness}
\label{sec_skewness}

At $T=0$, it is easy to compute $\delta\rho/\rho_b=\delta({\bf x},t)$ in second order perturbation theory and determine the growth of the skewness due to nonlinear gravitational clustering. This is a classical calculation in cosmology \cite{peebles} that we shall adapt to the present situation. For a cold overdamped gas, the equations of the problem are
\begin{equation}
\label{sk1}
\xi\frac{\partial\delta}{\partial t}=S_d G\rho_b \delta (1+\delta)+\frac{1}{a^2}\nabla\delta\cdot \nabla\phi,
\end{equation}
\begin{equation}
\label{sk2}
\Delta\phi=S_d G \rho_b \delta a^2.
\end{equation}
It is convenient to measure the evolution in terms of $a$ instead of $t$. Using Eq. (\ref{h8}), we obtain
\begin{equation}
\label{sk3}
-\frac{a}{d}\frac{\partial\delta}{\partial a}=\delta+\delta^2+\nabla\delta\cdot \nabla\chi,
\end{equation}
\begin{equation}
\label{sk4}
\Delta\chi=\delta, \qquad \chi=\frac{\phi}{S_d G\rho_b a^2}.
\end{equation}
In the linear approximation, the density contrast and the gravitational potential are given by
\begin{equation}
\label{sk5}
\delta_0({\bf x},a)=\frac{D({\bf x})}{a^d},\qquad \chi_0({\bf x},a)=\frac{C({\bf x})}{a^d},
\end{equation}
with $\Delta C=D$. To estimate the deviation to the linear regime, we write
\begin{equation}
\label{sk6}
\delta({\bf x},a)=\delta_0({\bf x},a)\lbrack 1+\epsilon({\bf x},a)\rbrack,
\end{equation}
with $\delta_0\ll 1$ and $\epsilon\ll 1$. Substituting Eq. (\ref{sk6}) in Eq. (\ref{sk3}), we obtain at leading order
\begin{equation}
\label{sk7}
-\frac{a}{d}\frac{\partial\epsilon}{\partial a}=\delta_0+\frac{\nabla\delta_0}{\delta_0}\cdot \nabla\chi_0.
\end{equation}
Using the expressions  of $\delta_0$ and $\chi_0$ given in Eq. (\ref{sk5}), the foregoing equation can be rewritten
\begin{equation}
\label{sk8}
\frac{\partial\epsilon}{\partial a}=-\frac{d}{a^{d+1}}\left (D+\frac{\nabla D}{D}\cdot \nabla C\right ).
\end{equation}
Integrating over $a$, we obtain
\begin{equation}
\label{sk9}
\epsilon=\frac{1}{a^{d}}\left (D+\frac{\nabla D}{D}\cdot \nabla C\right )=\delta_0+\frac{\nabla\delta_0}{\delta_0}\cdot \nabla\chi_0.
\end{equation}
Therefore, the expression of the density contrast in second order perturbation theory is
\begin{equation}
\label{sk10}
\delta=\delta_0+\delta_0^2+\nabla\delta_0\cdot \nabla\chi_0.
\end{equation}
This equation shows that, at this order, the behavior of the system  is non-local since the density contrast in ${\bf x}$ depends on the values of $\delta_0$ in ${\bf x}'$ through the function $\chi_0$ that is solution of the Poisson equation $\Delta\chi_0=\delta_0$.

An interesting application of the previous result concerns the growth of the skewness of the distribution $\delta({\bf x},t)$. We assume that the initial distribution $\delta_i({\bf x})$ is a random Gaussian process with zero mean and autocorrelation function $\xi(x)$. Therefore $\langle \delta_i\rangle=0$ and $\langle \delta_i^2\rangle=\xi(0)$. In the linear regime, the process remains Gaussian and the autocorrelation increases like $\xi(x,a)\propto 1/a^{2d}$. In the nonlinear regime, the relation $\langle \delta\rangle =0$ must be preserved because of mass conservation. This can be checked explicitly at second order from Eq. (\ref{sk10}). Indeed,
\begin{equation}
\label{sk11}
\langle \delta\rangle =\langle\delta_0\rangle+\langle \delta_0^2\rangle +\langle \nabla\delta_0\cdot \nabla\chi_0\rangle=\langle \delta_0^2\rangle -\langle \delta_0^2\rangle=0,
\end{equation}
where we have used an integration by parts, and the Poisson equation (\ref{sk4}), to get the second equality. The interesting moment is the skewness $\langle \delta^3\rangle$ since the initial distribution (Gaussian) has zero skewness. From Eqs. (\ref{sk6}) and (\ref{sk9}), we obtain at lowest order
\begin{eqnarray}
\label{sk12}
\langle \delta^3\rangle =\langle\delta_0^3(1+3\epsilon)\rangle=3\langle \delta_0^3\epsilon\rangle=3\langle \delta_0^4\rangle+3\langle\delta_0^2\nabla\delta_0\cdot\nabla\chi_0\rangle.\nonumber\\
\end{eqnarray}
This expression can be simplified as follows
\begin{equation}
\label{sk13}
\langle \delta^3\rangle=3\langle \delta_0^4\rangle+\langle\nabla\delta_0^3\cdot\nabla\chi_0\rangle=3\langle \delta_0^4\rangle-\langle\delta_0^4\rangle=2\langle \delta_0^4\rangle.
\end{equation}
For a Gaussian distribution, the Kurtosis is equal to $3$ i.e. $\langle\delta_0^4\rangle=3\xi(0)^2$. Therefore, the skewness of the density fluctuations is
\begin{equation}
\label{sk14}
\langle \delta^3\rangle=6\xi(0)^2,
\end{equation}
where, to lowest order, the variance is $\langle \delta^2\rangle=\xi(0)$. This leads to our final result
\begin{equation}
\label{sk15}
\frac{\langle \delta^3\rangle}{\langle\delta^2\rangle^2}=6.
\end{equation}

\subsection{An exact integral equation in Fourier space}
\label{sec_eie}

It is somewhat easier to analyze the nonlinear evolution of the system in Fourier space rather than in direct space. We shall therefore rewrite the exact equations (\ref{c12})-(\ref{c13}) for the density contrast in Fourier space. We follow a method similar to the one developed in cosmology \cite{peebles,paddycosmo}.

If we restrict ourselves to ideal self-gravitating Brownian particles described by an isothermal equation of state (\ref{sp3}), the fundamental equations of the problem are
\begin{equation}
\label{eie1}
\xi\frac{\partial\delta}{\partial t}=\frac{1}{a^2}\nabla\cdot \left\lbrack \frac{k_BT}{m}\nabla \delta+(1+\delta)\nabla\phi\right\rbrack,
\end{equation}
\begin{equation}
\label{eie2}
\Delta\phi=S_d G \rho_b \delta a^2.
\end{equation}
Let us decompose the density contrast and the gravitational potential in Fourier modes
\begin{equation}
\label{eie3}
\delta({\bf x},t)=\int \delta_{\bf k}(t)e^{i{\bf k}\cdot {\bf x}}\, \frac{d{\bf k}}{(2\pi)^d},
\end{equation}
\begin{equation}
\label{eie4}
\phi({\bf x},t)=\int \phi_{\bf k}(t)e^{i{\bf k}\cdot {\bf x}}\, \frac{d{\bf k}}{(2\pi)^d}.
\end{equation}
The Poisson equation (\ref{eie2}) implies
\begin{eqnarray}
\label{eie5}
\delta_{\bf k}=-\frac{k^2\phi_{\bf k}}{S_d G \rho_b a^2}=-\frac{k^2 a^{d-2}\phi_{\bf k}}{S_d G \rho_b a^d},
\end{eqnarray}
where we recall that $\rho_b a^d=dM/S_d$ is constant. On the other hand, substituting Eqs. (\ref{eie3}) and (\ref{eie4}) in Eq. (\ref{eie1}), and using Eq. (\ref{eie5}), we obtain after simple calculations
\begin{eqnarray}
\label{eie6}
\xi{\dot \delta}_{\bf k}=-\frac{k_BT}{ma^2}k^2\delta_{\bf k}+S_d G\rho_b\delta_{\bf k}\nonumber\\
+S_dG\rho_b\int \delta_{{\bf k}'}\delta_{{\bf k}-{\bf k}'}\frac{{\bf k}\cdot {\bf k}'}{{k'}^2}\, \frac{d{\bf k}'}{(2\pi)^d}.
\end{eqnarray}
Symmetrizing the last term, we finally get
\begin{eqnarray}
\label{eie7}
\xi{\dot \delta}_{\bf k}=-\frac{k_BT}{ma^2}k^2\delta_{\bf k}+S_d G\rho_b\delta_{\bf k}\nonumber\\
+\frac{S_dG\rho_b}{2}\int \delta_{{\bf k}'}\delta_{{\bf k}-{\bf k}'}\left\lbrack \frac{{\bf k}\cdot {\bf k}'}{{k'}^2}+\frac{{\bf k}\cdot ({\bf k}-{\bf k}')}{|{\bf k}-{\bf k'}|^2} \right\rbrack \, \frac{d{\bf k}'}{(2\pi)^d}.
\end{eqnarray}
Using Eq. (\ref{eie5}), we can obtain an equation for the Fourier components of the gravitational potential. After simple calculations and re-arrangement of terms, it can be written
\begin{eqnarray}
\label{eie8}
\xi{\dot \phi}_{\bf k}+2(d-1)\xi\frac{\dot a}{a}\phi_{\bf k}=-\frac{k_BT}{ma^2}k^2\phi_{\bf k}
-\frac{1}{2a^2}\nonumber\\
\times\int \phi_{\frac{1}{2}{\bf k}+{\bf p}}\phi_{\frac{1}{2}{\bf k}-{\bf p}}\left\lbrack \left (\frac{k}{2}\right )^2+p^2-2\left (\frac{{\bf k}\cdot {\bf p}}{k}\right )^2\right\rbrack\, \frac{d{\bf p}}{(2\pi)^d},\nonumber\\
\end{eqnarray}
where we have defined ${\bf p}={\bf k}'-{\bf k}/2$. Measuring the evolution in terms of $a$ instead of $t$, and using Eq. (\ref{h8}), we obtain the equivalent equations
\begin{eqnarray}
\label{eie9}
\frac{a}{d}\frac{d{\delta}_{\bf k}}{da}=\left (\frac{k_BTk^2}{S_d G\rho_b ma^2}-1\right )\delta_{\bf k}\nonumber\\
-\frac{1}{2}\int \delta_{{\bf k}'}\delta_{{\bf k}-{\bf k}'}\left\lbrack \frac{{\bf k}\cdot {\bf k}'}{{k'}^2}+\frac{{\bf k}\cdot ({\bf k}-{\bf k}')}{|{\bf k}-{\bf k'}|^2} \right\rbrack \, \frac{d{\bf k}'}{(2\pi)^d},
\end{eqnarray}
and
\begin{eqnarray}
\label{eie10}
a\frac{d{\phi}_{\bf k}}{da}+2(d-1)\phi_{\bf k}=\frac{k_BTa^{d-2}}{GMm}k^2\phi_{\bf k}
+\frac{a^{d-2}}{2GM}\nonumber\\
\times\int \phi_{\frac{1}{2}{\bf k}+{\bf p}}\phi_{\frac{1}{2}{\bf k}-{\bf p}}\left\lbrack \left (\frac{k}{2}\right )^2+p^2-2\left (\frac{{\bf k}\cdot {\bf p}}{k}\right )^2\right\rbrack\, \frac{d{\bf p}}{(2\pi)^d}.\nonumber\\
\end{eqnarray}
We stress that these equations are exact and {\it closed} while the equivalent equations in cosmology \cite{peebles,paddycosmo} are not closed.  These equations, however, like the original Smoluchowski-Poisson system (\ref{sp1})-(\ref{sp2}), rely on a meanfield approximation. More exact (but less useful) equations, which do not make the meanfield approximation, are derived in Appendix \ref{sec_der} for $T=0$.

Like in cosmology, it is possible to derive interesting results from these integral equations in Fourier space. We shall mention in particular the $k^4$ tail and the self-similar solution.

\subsubsection{Inverse cascade: the $k^4$ tail}
\label{sec_tail}

We consider the case of a cold overdamped gas ($T=0$). In the linear regime, the density contrast and the gravitational potential are given by
\begin{eqnarray}
\label{eie11}
\delta^{(L)}({\bf x},a)=\frac{D({\bf x})}{a^d}, \quad \phi^{(L)}({\bf x},a)=\frac{S_d G\rho_b a^d Q({\bf x})}{a^{2(d-1)}},
\end{eqnarray}
where $Q$ is solution of the Poisson equation
\begin{eqnarray}
\label{eie12}
\Delta Q=D,\qquad Q_{\bf k}=-\frac{D_{\bf k}}{k^2}.
\end{eqnarray}
Let us suppose that the initial power spectrum for the density has very little power at large scales so that $P_\delta(k)\propto k^n$ with $n>4$ for small $k$. In the linear regime, the power spectrum evolves like $P_\delta(k)\propto k^n/a^{2d}$ so that its shape in $k$-space is not modified and is always subdominant to $k^4$. However, when nonlinear effects come into play, a ``$k^4$ tail'' develops for $k\rightarrow 0$ (see below) and the contribution of the large-scale structures finally dominate the power spectrum. This result was first found in cosmology \cite{peebles,paddycosmo}  and can be interpreted as a sort of ``inverse cascade'' like in two-dimensional turbulence. In 2D turbulence, the inverse cascade is associated with the formation of large-scale vortices. Similarly, in the present problem, it is associated with the formation of virialized clusters (in cosmology, these clusters correspond to galaxies and dark matter halos). It may be noted that the equations of 2D turbulence for the coarse-grained vorticity field \cite{rs,csr} are relatively similar to the Smoluchowski-Poisson system. This remark can reinforce the analogy between self-gravitating (Brownian) systems and two-dimensional turbulence \cite{houches}.

To see the emergence of the $k^4$ tail in  nonlinear gravitational clustering, we write $\delta=\delta^{(L)}+\delta^{(2)}+...$ where $\delta^{(1)}=\delta^{(L)}$ is the density contrast in the linear theory and $\delta^{(2)}$ is the next order correction. To second order, we get from Eq. (\ref{eie9}) with $T=0$ the equation
\begin{eqnarray}
\label{eie13}
\frac{a}{d}\frac{d{\delta}^{(2)}_{\bf k}}{da}+\delta^{(2)}_{\bf k}=\qquad\qquad\qquad\qquad\nonumber\\
-\frac{1}{2}\int \delta^{(L)}_{{\bf k}'}\delta^{(L)}_{{\bf k}-{\bf k}'}\left\lbrack \frac{{\bf k}\cdot {\bf k}'}{{k'}^2}+\frac{{\bf k}\cdot ({\bf k}-{\bf k}')}{|{\bf k}-{\bf k'}|^2} \right\rbrack \, \frac{d{\bf k}'}{(2\pi)^d}.
\end{eqnarray}
Using Eq. (\ref{eie11}), it can be rewritten
\begin{eqnarray}
\label{eie14}
\frac{a}{d}\frac{d{\delta}^{(2)}_{\bf k}}{da}+\delta^{(2)}_{\bf k}=\qquad\qquad\qquad\qquad\nonumber\\
-\frac{1}{2a^{2d}}\int D_{{\bf k}'}D_{{\bf k}-{\bf k}'}\left\lbrack \frac{{\bf k}\cdot {\bf k}'}{{k'}^2}+\frac{{\bf k}\cdot ({\bf k}-{\bf k}')}{|{\bf k}-{\bf k'}|^2} \right\rbrack \, \frac{d{\bf k}'}{(2\pi)^d}.
\end{eqnarray}
The solution to this equation is the sum of the solution  of the homogeneous part (which increases like $1/a^{d}$) and a particular solution (which increases like ${1}/a^{2d}$). We shall only keep the particular solution which dominates for large times. Writing $\delta_{\bf k}=C_{\bf k}/a^{2d}$, determining the constant $C_{\bf k}$ from Eq. (\ref{eie14}), and reintroducing the original variables, we obtain the solution
\begin{eqnarray}
\label{eie15}
{\delta}^{(2)}_{\bf k}=\frac{1}{2}\int \delta^{(L)}_{{\bf k}'}\delta^{(L)}_{{\bf k}-{\bf k}'}\left\lbrack \frac{{\bf k}\cdot {\bf k}'}{{k'}^2}+\frac{{\bf k}\cdot ({\bf k}-{\bf k}')}{|{\bf k}-{\bf k'}|^2} \right\rbrack \, \frac{d{\bf k}'}{(2\pi)^d}.
\end{eqnarray}
For $k\rightarrow 0$, we get
\begin{eqnarray}
\label{eie16}
{\delta}^{(2)}_{{\bf k}\simeq {\bf 0}}\sim \frac{1}{2}k^2\int \frac{|\delta^{(L)}_{{\bf k}'}|^2}{{k'}^2} \, \frac{d{\bf k}'}{(2\pi)^d}\propto \frac{k^2}{a^{2d}}.
\end{eqnarray}
This implies that the power spectrum for the density behaves like $P_{\delta}(k)\propto k^4/a^{4d}$ for $k\rightarrow 0$.

Of course, the same result can be obtained from Eq. (\ref{eie10}). Writing $\phi=\phi^{(L)}+\phi^{(2)}+...$ and using Eq. (\ref{eie11}),  we obtain
\begin{eqnarray}
\label{eie17}
a\frac{d{\phi}^{(2)}_{\bf k}}{da}+2(d-1)\phi^{(2)}_{\bf k}\nonumber\\
=\frac{a^{d-2}}{2GM}\int \phi^{(L)}_{\frac{1}{2}{\bf k}+{\bf p}}\phi^{(L)}_{\frac{1}{2}{\bf k}-{\bf p}}G({\bf k},{\bf p})\, \frac{d{\bf p}}{(2\pi)^d}\nonumber\\
=\frac{(S_d G\rho_b a^d)^2}{2GM a^{3d-2}}\int Q_{\frac{1}{2}{\bf k}+{\bf p}}Q_{\frac{1}{2}{\bf k}-{\bf p}}G({\bf k},{\bf p})\, \frac{d{\bf p}}{(2\pi)^d}.
\end{eqnarray}
Proceeding as before, the solution to this equation is
\begin{eqnarray}
\label{eie18}
\phi^{(2)}_{\bf k}=-\frac{a^{d-2}}{2dGM}\int \phi^{(L)}_{\frac{1}{2}{\bf k}+{\bf p}}\phi^{(L)}_{\frac{1}{2}{\bf k}-{\bf p}}G({\bf k},{\bf p})\, \frac{d{\bf p}}{(2\pi)^d}.
\end{eqnarray}
For $k\rightarrow 0$, we get
\begin{eqnarray}
\label{eie19}
\phi^{(2)}_{{\bf k}\simeq {\bf 0}}\simeq -\frac{a^{d-2}}{2dGM}\int |\phi^{(L)}_{{\bf p}}|^2 p^2\, \frac{d{\bf p}}{(2\pi)^d}\propto \frac{1}{a^{3d-2}},
\end{eqnarray}
which is independent on ${\bf k}$. Using Eqs. (\ref{eie5}) and (\ref{eie19}), the power spectrum for the density behaves like $P_{\delta}(k)\sim k^4 a^{2(d-2)}P_{\phi}(k)\propto k^4/a^{4d}$ for $k\rightarrow 0$.

\subsubsection{Self-similar solution in the fully nonlinear regime}
\label{sec_sel}

At $T=0$, Eq. (\ref{eie9}) admits self-similar solutions of the form $\delta_{\bf k}(a)=\delta(a)D_{\bf k}$. This self-similar solution is expected to be reached for sufficiently large times when the initial condition becomes irrelevant. This implies that nonlinear gravitational clustering leads to a universal power spectrum at late times. Again, this result is similar to the one obtained in cosmology \cite{paddycosmo}. However, in the present case, the self-similar solution is {\it exact} while in cosmology it is approximate (although very accurate) since the Zeldovich approximation must be used to close the integral equation for the density contrast.

Substituting the ansatz  $\delta_{\bf k}(a)=\delta(a)D_{\bf k}$ in Eq. (\ref{eie9}), we obtain the separate equations
\begin{eqnarray}
\label{eie20}
\frac{a}{d}\frac{d{\delta}}{da}+\delta=\mu\delta^2,
\end{eqnarray}
\begin{eqnarray}
\label{eie21}
\mu D_{\bf k}=-\frac{1}{2}\int D_{{\bf k}'}D_{{\bf k}-{\bf k}'}\left\lbrack \frac{{\bf k}\cdot {\bf k}'}{{k'}^2}+\frac{{\bf k}\cdot ({\bf k}-{\bf k}')}{|{\bf k}-{\bf k'}|^2} \right\rbrack \, \frac{d{\bf k}'}{(2\pi)^d},\nonumber\\
\end{eqnarray}
where $\mu$ is an arbitrary constant.  Eq. (\ref{eie20}) can be easily solved yielding
\begin{eqnarray}
\label{eie22}
\delta(a)=\frac{1}{\mu-(\frac{a}{a_0})^d\lbrack\mu-\frac{1}{\delta(a_0)}\rbrack}.
\end{eqnarray}
We note that $\delta(a)\rightarrow 1/\mu$ for large times ($a\rightarrow 0$) showing that nonlinear effects finally stabilize the system. Equations (\ref{eie21}) and (\ref{eie22}) provide  an {\it exact} solution of the nonlinear problem.  For $k\rightarrow 0$, Eq. (\ref{eie21}) reduces to
\begin{eqnarray}
\label{eie23}
\mu D_{{\bf k}\simeq {\bf 0}}\sim -\frac{1}{2}k^2\int \frac{|D_{{\bf k}'}|^2}{{k'}^2}\, \frac{d{\bf k}'}{(2\pi)^d},
\end{eqnarray}
implying that the power spectrum for the density behaves like $P_{\delta}(k)\propto \delta(a)^2k^4$ for $k\rightarrow 0$.

Of course, we can do the same analysis from Eq. (\ref{eie10}). Substituting $\phi_{\bf k}(a)=F(a)S_d G\rho_b a^d Q_{\bf k}$ in Eq. (\ref{eie10}), we get the separate equations
\begin{eqnarray}
\label{eie24}
a\frac{dF}{da}+2(d-1)F=d\mu a^{d-2} F^2,
\end{eqnarray}
\begin{eqnarray}
\label{eie25}
\mu Q_{\bf k}=\frac{1}{2}\int Q_{\frac{1}{2}{\bf k}+{\bf p}}Q_{\frac{1}{2}{\bf k}-{\bf p}}G({\bf k},{\bf p})\, \frac{d{\bf p}}{(2\pi)^d},
\end{eqnarray}
with $G({\bf k},{\bf p})=({k}/{2})^2+p^2-2({{\bf k}\cdot {\bf p}}/{k})^2$. This is consistent with the preceding formulae if we set $\delta(a)=F(a)a^{d-2}$ and $D_{\bf k}=-k^2Q_{\bf k}$. For $k\rightarrow 0$, Eq. (\ref{eie25}) reduces to
\begin{eqnarray}
\label{eie26}
\mu Q_{{\bf k}\simeq {\bf 0}}=\frac{1}{2}\int |Q_{{\bf p}}|^2 p^2\, \frac{d{\bf p}}{(2\pi)^d},
\end{eqnarray}
implying that the power spectrum for the density behaves like $P_{\delta}(k)\sim k^4 a^{2(d-2)}P_{\phi}(k)\propto F(a)^2 a^{2(d-2)} k^4$ for $k\rightarrow 0$.

{\it Remark:} in Appendix \ref{sec_sss}, we show that the self-similar solution can be directly obtained in physical space (instead of Fourier space).

\subsection{Spherical approximation}
\label{sec_sph}

In the nonlinear regime, it is not possible to analytically solve Eqs. (\ref{c12})-(\ref{c13}) in full generality. We can, however, obtain a particular solution if we assume that, locally, the density is homogeneous. Therefore, we shall describe the evolution of a constant overdensity compared with the background. This is the equivalent of the {\it spherical top hat} in cosmology \cite{peebles,paddycosmo}. If we assume $\delta({\bf x},t)=\delta(t)$, Eq. (\ref{c12}) becomes
\begin{equation}
\label{sph1}
\xi\frac{d\delta}{dt}=S_d G\rho_b \delta (1+\delta).
\end{equation}
Let us assume that this equation describes the evolution of a spherical region of mass ${\cal M}$ and radius $R(t)$. In that case, the density contrast can be written
\begin{equation}
\label{sph2}
1+\delta=\frac{\rho}{\rho_b}=\frac{{\cal M}}{M}\left (\frac{a}{R}\right )^d.
\end{equation}
Substituting Eq. (\ref{sph2}) in Eq. (\ref{sph1}) and using Eq. (\ref{h8}), we obtain
\begin{equation}
\label{sph3}
\xi\frac{dR}{dt}=-\frac{G{\cal M}}{R^{d-1}},\quad {\rm or}\quad \xi\frac{d\rho}{dt}=S_d G\rho^2.
\end{equation}
This corresponds to the collapse of a uniform sphere described by the BSP system (\ref{sp1})-(\ref{sp2}).

This observation leads to the following scenario. If we start from a uniform sphere of mass $M$ and radius $a_0$, this sphere is unstable and collapses under its own gravity. The mean flow develops a finite time singularity: the average density $\rho_b(t)$ becomes infinite in a finite time $t_{*}$ while the radius $a(t)$ of the sphere tends to zero. However, this sphere is unstable to small perturbations and a process of fragmentation follows. The details on this process depend on the equation of state, e.g. on the polytropic index. If $\gamma>\gamma_{4/3}$, the system is asymptotically stable. However, the perturbations whose wavelengths $\lambda$ are larger than the initial Jeans wavelength $\lambda_J(0)$ grow (see Sec. \ref{sec_sol}). This initial growth can trigger nonlinear effects and result in the formation of overdense regions. At first, these regions experience a cold collapse and intensify. Then, when pressure effects become important at high densities and small scales, localized clusters in virial equilibrium form. These individual clusters, which correspond to complete polytropic spheres, are stable since $\gamma>\gamma_{4/3}$.  If $\gamma<\gamma_{4/3}$, the collapsing sphere of radius $a(t)$ is unstable to small perturbations whatever their wavelength. In the linear regime, the density contrast is small  ($\delta\ll 1$) and the perturbations grow as described in Sec. \ref{sec_sol}. In the nonlinear regime, the density contrast becomes large ($\delta\gg 1$) and the system forms overdense regions. They correspond to  localized clusters. However, these individual clusters are unstable since $\gamma<\gamma_{4/3}$ and they experience gravitational collapse. They can collapse as uniforms spheres and fragmentate into smaller pieces as described in this paper. This leads to a process of hierarchical clustering in which clusters fragmentate into smaller clusters which themselves fragmentate into smaller clusters and so on. Alternatively, the clusters can undergo a self-similar spherically symmetric collapse up to a Dirac peak as described in \cite{crs,sciso,post,virial1,cspoly1,scpoly2}. In that case, there is no fragmentation. The dynamics of these clusters is also interesting since they interact with each other. At large scales, their internal structure can be neglected and the problem is reduced to the dynamical evolution of a system of ${\cal N}$ clusters in interaction. When two clusters come into contact, they merge so that their mass ${\cal M}(t)$ increases while their number ${\cal N}(t)$ decreases. Ultimately, only one cluster remains: a Dirac peak containing all the mass (statistical equilibrium state in the canonical ensemble). This is reminiscent of a coarsening process in statistical mechanics \cite{bray}. This also shares analogies with the process of decaying two-dimensional turbulence \cite{pomeau}. Starting from an incoherent initial condition, large-scale vortices spontaneously emerge  in a 2D incompressible flow and rapidly dominate the dynamics. These vortices evolve under the effect of their mutual advection (due to long-range interactions) punctuated by merging events. Their number decreases as a power law $N(t)\sim t^{-\xi}$  with an exponent $\xi\simeq 0.6-1$ \cite{scvortex}. It would be interesting to further develop the analogy between 2D decaying turbulence and self-gravitating Brownian particles.

\section{Quantum Smoluchowski-Poisson system}
\label{sec_qsp}

We now introduce and study a more general Smoluchowski equation taking  quantum effects into account. To that purpose, we consider a dissipative gas of self-gravitating Bose-Einstein condensates. Using the Madelung transformation, we show that the dissipative Gross-Pitaevskii equation is equivalent to the damped barotropic Euler equations with an additional quantum potential. In a strong friction limit, this yields the quantum barotropic Smoluchowski equation.  We extend our previous analysis to this more general equation.

\subsection{Dissipative Gross-Pitaevskii-Poisson system}
\label{sec_mfgp}

We consider an interacting gas of Bose-Einstein condensates at $T=0$ described by the dissipative mean field Gross-Pitaevskii (GP) equation
\cite{gross,pitaevskii}:
\begin{eqnarray}
\label{mfgp1}
i\hbar \frac{\partial\psi}{\partial t}({\bf r},t)=-\frac{\hbar^2}{2m}\Delta\psi({\bf r},t)+m\Phi_{tot}({\bf r},t)\psi({\bf r},t)\nonumber\\
+\frac{\xi\hbar}{2i}\ln\left (\frac{\psi}{\psi^*}\right )\psi,
\end{eqnarray}
\begin{eqnarray}
\label{mfgp2}
\Phi_{tot}({\bf r},t)=\int \rho({\bf r}',t) u(|{\bf r}-{\bf r}'|)\, d{\bf r}',
\end{eqnarray}
\begin{eqnarray}
\label{mfgp3}
\rho({\bf r},t)=Nm|\psi({\bf r},t)|^2,
\end{eqnarray}
\begin{eqnarray}
\label{mfgp4}
\int |\psi({\bf r},t)|^2\, d{\bf r}=1.
\end{eqnarray}
Equation (\ref{mfgp4}) is the normalization condition, Eq. (\ref{mfgp3}) gives the density of the BEC, Eq. (\ref{mfgp2}) determines the associated potential and Eq. (\ref{mfgp1}) determines the wave function. We assume that the potential of interaction can be written as $u=u_{LR}+u_{SR}$ where $u_{LR}$ refers to the long-range gravitational interaction and $u_{SR}$ to the short-range interaction. We assume that the short-range interaction corresponds to binary collisions that can be modeled by the effective potential $u_{SR}({\bf r}-{\bf r}')=g\delta({\bf r}-{\bf r}')$, where the coupling constant (or pseudo-potential) $g$ is related to the $s$-wave scattering length $a_s$ through $g=4\pi a_s\hbar^2/m^3$ (in $d=3$) \cite{revuebec}. For the sake of generality, we allow $a_s$ to be positive or negative ($a_s>0$ corresponds to short-range repulsions and $a_s<0$ corresponds to short-range attractions). Under these conditions, the total potential can be written $\Phi_{tot}=\Phi+h(\rho)$ where $\Phi$ is the gravitational potential and
\begin{equation}
\label{mfgp5}
h(\rho)=g\rho=gNm|\psi|^2,
\end{equation}
is an effective potential modeling short-range interactions.  For the sake of generality, we will consider an arbitrary function $h(\rho)$. We shall also assume that the particles are subjected to an external potential $\Phi_{ext}({\bf r})$. For illustration, we may consider a harmonic potential $\Phi_{ext}=\omega_0^2 r^2/2$ which can play the role of a trap ($\omega_0^2>0$) or a rotation ($\omega_0^2=-\Omega^2<0$). Regrouping these results, we obtain the dissipative  Gross-Pitaevskii-Poisson (GPP) system
\begin{eqnarray}
\label{mfgp6}
i\hbar \frac{\partial\psi}{\partial t}({\bf r},t)=-\frac{\hbar^2}{2m}\Delta\psi({\bf r},t)
+m\lbrack\Phi({\bf r},t)\nonumber\\
+h(\rho)+\Phi_{ext}({\bf r})\rbrack\psi({\bf r},t)+\frac{\xi\hbar}{2i}\ln\left (\frac{\psi}{\psi^*}\right )\psi,
\end{eqnarray}
\begin{equation}
\label{mfgp7}
\Delta\Phi=S_d G\rho=S_d G Nm |\psi|^2.
\end{equation}
Instead of the gravitational potential, we could consider an arbitrary binary potential with long-range interactions of the form $u_{LR}(|{\bf r}-{\bf r}'|)$ in which case the Poisson equation would be replaced by $\Phi=u_{LR}*\rho$ where $*$ denotes the product of convolution.

We have  introduced a source of dissipation in the Gross-Pitaevskii equation measured by the coefficient $\xi$ (the ordinary Gross-Pitaevskii equation is recovered for $\xi=0$). Initially, we ``guessed'' the form of this dissipation term in order to obtain, after making the Madelung transformation, the damped Euler equations (as we shall see, the parameter $\xi$ can be interpreted as a friction coefficient in the Euler equation (\ref{mad10})). Then, we found some confidence in the fact that the same term was previously derived by Kostin \cite{kostin} from the Heisenberg-Langevin equation describing a Brownian particle interacting with a thermal bath environment \footnote{In Kostin's approach, the Gross-Pitaevskii equation also contains a random potential. In our approach, this stochastic term is neglected since we assumed $T=0$.}. We can therefore consider that Eqs. (\ref{mfgp6})-(\ref{mfgp7}) have some physical foundation. Recently, some authors \cite{bohmer} have proposed that dark matter in the universe could be a self-gravitating Bose-Einstein condensate described by the GPP system (\ref{mfgp6})-(\ref{mfgp7}) with $\xi=0$ (see \cite{paper1} for a short history). Here, we consider a more general model where damping effects are taken into account. In addition to cosmology, this model can probably find applications in other domains of physics.

\subsection{Madelung transformation}
\label{sec_mad}

Let use the Madelung \cite{madelung} transformation to rewrite the dissipative GPP system in the form of hydrodynamic equations. We write the wavefunction in the form
\begin{equation}
\label{mad1}
\psi({\bf r},t)=A({\bf r},t) e^{iS({\bf r},t)/\hbar}
\end{equation}
where $A({\bf r},t)$ and $S({\bf r},t)$ are real functions. We clearly have
\begin{equation}
\label{mad2}
A=\sqrt{|\psi|^2},\qquad S=\frac{\hbar}{2i}\ln\left (\frac{\psi}{\psi^*}\right ).
\end{equation}
Note that the dissipative term in the GP equation (\ref{mfgp6}) can be written $\xi S\psi$. Substituting Eq. (\ref{mad1}) in Eq. (\ref{mfgp6}) and separating real and imaginary parts, we obtain
\begin{equation}
\label{mad3}
\frac{\partial S}{\partial t}+\frac{1}{2m}(\nabla S)^2+m\Phi_{tot}-\frac{\hbar^2}{2m}\frac{\Delta A}{A}+\xi S=0,
\end{equation}
\begin{equation}
\label{mad4}
\frac{\partial A^2}{\partial t}+\nabla\cdot \left (\frac{A^2 \nabla S}{m}\right )=0.
\end{equation}
Following Madelung, we introduce the density and the velocity fields
\begin{equation}
\label{mad5}
\rho=NmA^2=Nm|\psi|^2, \qquad  {\bf u}=\frac{1}{m}\nabla S.
\end{equation}
We first note that the flow is irrotational since $\nabla\times {\bf u}={\bf 0}$.  With these variables, Eqs. (\ref{mad3}) and (\ref{mad4}) can be rewritten
\begin{equation}
\label{mad6}
\frac{\partial\rho}{\partial t}+\nabla\cdot (\rho {\bf u})=0,
\end{equation}
\begin{equation}
\label{mad7}
\frac{\partial S}{\partial t}+\frac{1}{2m}(\nabla S)^2+m(\Phi+h(\rho)+\Phi_{ext})+Q+\xi S=0,
\end{equation}
where
\begin{equation}
\label{mad8}
Q=-\frac{\hbar^2}{2m}\frac{\Delta \sqrt{\rho}}{\sqrt{\rho}}=-\frac{\hbar^2}{4m}\left\lbrack \frac{\Delta\rho}{\rho}-\frac{1}{2}\frac{(\nabla\rho)^2}{\rho^2}\right\rbrack,
\end{equation}
is the quantum potential.  The first equation is similar to the equation of continuity in hydrodynamics. The second equation has a form similar to the classical Hamilton-Jacobi equation with an additional quantum term and a source of dissipation. It  can also be interpreted as a generalized Bernouilli equation for a potential flow. Taking the gradient of Eq. (\ref{mad7}), and using the well-known  identity $({\bf u}\cdot \nabla){\bf u}=\nabla ({{\bf u}^2}/{2})-{\bf u}\times (\nabla\times {\bf u})$ which reduces to $({\bf u}\cdot \nabla){\bf u}=\nabla ({{\bf u}^2}/{2})$ for an irrotational flow, we obtain an equation similar to the damped Euler equation with an additional quantum potential
\begin{equation}
\label{mad9}
\frac{\partial {\bf u}}{\partial t}+({\bf u}\cdot \nabla){\bf u}=-\nabla h-\nabla\Phi-\nabla \Phi_{ext}-\frac{1}{m}\nabla Q-\xi {\bf u}.
\end{equation}
This equation shows that the effective potential $h$ appearing in the GP equation can be
interpreted as an enthalpy in the hydrodynamic equations.  We can rewrite Eq. (\ref{mad9}) in the form
\begin{equation}
\label{mad10}
\frac{\partial {\bf u}}{\partial t}+({\bf u}\cdot \nabla){\bf u}=-\frac{1}{\rho}\nabla p-\nabla\Phi-\nabla \Phi_{ext}-\frac{1}{m}\nabla Q-\xi {\bf u},
\end{equation}
where $p({\bf r},t)$ is a pressure.  Since $h({\bf r},t)=h\lbrack \rho({\bf r},t)\rbrack$, the pressure $p({\bf r},t)=p\lbrack \rho({\bf r},t)\rbrack$ is a function of the density  (the flow is barotropic). The equation of state $p(\rho)$ is determined by the potential $h(\rho)$ through the relation
\begin{equation}
\label{mad11}
h'(\rho)=\frac{p'(\rho)}{\rho}.
\end{equation}
This yields $p(\rho)=\rho h(\rho)-H(\rho)$ where $H$ is a primitive of $h$. In conclusion, the dissipative GPP system is {\it equivalent} to the hydrodynamic equations
\begin{equation}
\label{mad12}
\frac{\partial\rho}{\partial t}+\nabla\cdot (\rho {\bf u})=0,
\end{equation}
\begin{equation}
\label{mad13}
\frac{\partial {\bf u}}{\partial t}+({\bf u}\cdot \nabla){\bf u}=-\frac{1}{\rho}\nabla p-\nabla\Phi-\nabla \Phi_{ext}-\frac{1}{m}\nabla Q-\xi{\bf u},
\end{equation}
\begin{equation}
\label{mad14}
\Delta\Phi=S_d G\rho.
\end{equation}
For a harmonic potential, $\nabla\Phi_{ext}=\omega_0^2{\bf r}$. We shall refer to these equations as the quantum damped barotropic Euler-Poisson system. We note the identity
\begin{equation}
\label{mad11new}
-\frac{1}{m}\nabla Q\equiv -\frac{1}{\rho}\partial_j P_{ij},
\end{equation}
where $P_{ij}$ is the quantum stress (or pressure) tensor
\begin{equation}
\label{mad12new}
P_{ij}=-\frac{\hbar^2}{4m^2}\rho\, \partial_i\partial_j\ln\rho,
\end{equation}
or
\begin{equation}
\label{mad13new}
P_{ij}=\frac{\hbar^2}{4m^2}\left (\frac{1}{\rho}\partial_i\rho\partial_j\rho-\delta_{ij}\Delta\rho\right ).
\end{equation}
This shows that the quantum potential is equivalent to an anisotropic pressure.
  In the classical limit $\hbar\rightarrow 0$, the quantum potential disappears and we recover the ordinary damped barotropic Euler-Poisson system.

In the strong friction limit $\xi\rightarrow +\infty$, we can formally neglect the inertia of the particles in Eq. (\ref{mad13}) and obtain
\begin{equation}
\label{mad15}
\xi{\bf u}\simeq -\frac{1}{\rho}\nabla p-\nabla\Phi-\nabla \Phi_{ext}-\frac{1}{m}\nabla Q.
\end{equation}
Substituting this relation in the continuity equation (\ref{mad12}), we finally arrive at the quantum barotropic Smoluchowski-Poisson system
\begin{equation}
\label{mad16}
\xi\frac{\partial\rho}{\partial t}=\nabla\cdot\left (\nabla p+\rho\nabla\Phi+\rho\nabla \Phi_{ext}+\frac{\rho}{m}\nabla Q\right ),
\end{equation}
\begin{equation}
\label{mad17}
\Delta\Phi=S_d G \rho.
\end{equation}
In the classical limit $\hbar\rightarrow 0$, the quantum potential disappears and we recover the classical barotropic Smoluchowski-Poisson system (\ref{sp1})-(\ref{sp2}). We note that, due to the complex nature of the wave function, it is not possible to take the strong friction limit directly in the dissipative GP equation (\ref{mfgp6}). We must necessarily split this equation into its real and imaginary parts, which is achieved by means of the Madelung transformation, and take the limit $\xi\rightarrow +\infty$ in the Euler equation (\ref{mad13}) corresponding to the real part of the GP equation.

{\it Remark:} a system of classical Brownian particles with long and short range  interactions \cite{physicaA} is described by the generalized Smoluchowski equation
\begin{equation}
\label{ddft}
\xi\frac{\partial\rho}{\partial t}=\nabla\cdot\left (\frac{k_{B}T}{m}\nabla\rho+\nabla p_{ex}+\rho\nabla\Phi+\rho\nabla \Phi_{ext}\right ),
\end{equation}
where $p_{id}=\rho k_BT/m$ is the ideal kinetic pressure and $p_{ex}$ is the excess pressure taking into account short-range interactions. The total pressure is $p=p_{id}+p_{ex}$. In the present case, $T=0$, so that the pressure in Eq. (\ref{mad16}) corresponds to $p_{ex}$. Contrary to the kinetic pressure $p_{id}$, the pressure $p_{ex}$ arising from short-range interactions can be positive or negative (see Sec. \ref{sec_eos}). In the DDFT, this pressure is related to the excess free energy $F_{ex}[\rho]$ by $\nabla p_{ex}=\rho\nabla \delta F_{ex}/\delta\rho$ \cite{physicaA}. In the present case, according to Eq. (\ref{mad11}), we have $\nabla p_{ex}=\rho\nabla h$. Therefore, the excess free energy is $F_{ex}=\int H(\rho)\, d{\bf r}$ where $H$ is a primitive of $h$. Similarly, the quantum pressure in Eq. (\ref{mad16}) can be obtained from the quantum excess free energy $F_{ex}^Q=({1}/{m})\int \rho Q\, d{\bf r}$ via $(\rho/m)\nabla Q=\rho\nabla \delta F_{ex}^Q/\delta\rho$ (see Appendix \ref{sec_ht}).

\subsection{Equation of state}
\label{sec_eos}

The effective potential $h(\rho)$, which  takes into account short-range interactions (collisions) between particles, determines a barotropic equation of state $p(\rho)$ through the relation (\ref{mad11}). Inversely, specifying an equation of state $p=p(\rho)$, we can obtain the corresponding function $h(\rho)$. Let us give specific examples.

The isothermal equation of state $p=\rho k_BT_{eff}/m$ leads  to an effective  potential of the form
\begin{equation}
\label{eos1}
h(\rho)=\frac{k_B T_{eff}}{m}\ln\rho.
\end{equation}
The corresponding GP equation can be written
\begin{equation}
\label{eos2}
i\hbar \frac{\partial\psi}{\partial t}=-\frac{\hbar^2}{2m}\Delta\psi+(m\Phi+{2k_B T_{eff}}\ln|\psi|)\psi.
\end{equation}
Interestingly, we note that a nonlinear Schr\"odinger equation with a logarithmic potential similar to Eq. (\ref{eos2}) has been introduced long ago by Bialynicki \& Mycielski \cite{bialynicki} as a possible generalization of the Schr\"odinger equation in quantum mechanics.

The polytropic equation of state $p=K\rho^{\gamma}$ with $\gamma=1+1/n$ leads to an effective potential of the form
\begin{equation}
\label{eos3}
h(\rho)=\frac{K\gamma}{\gamma-1}\rho^{\gamma-1}.
\end{equation}
The corresponding GP equation can be written
\begin{equation}
\label{eos4}
i\hbar \frac{\partial\psi}{\partial t}=-\frac{\hbar^2}{2m}\Delta\psi+m(\Phi+\kappa |\psi|^{2/n})\psi,
\end{equation}
where $\kappa=K(n+1)(Nm)^{1/n}$. This is the usual general form of the GP equation considered in the literature \cite{sulem}. We may recall that classical and ultra-relativistic fermion stars at $T=0$  are equivalent to polytropes with index $n=n_{3/2}\equiv d/2$ and $n=n'_3\equiv d$ respectively \cite{wdsd}. The GPP system (\ref{eos4})-(\ref{mfgp7})  with $n=3/2$ in $d=3$, describing self-gravitating fermions beyond the Thomas-Fermi approximation (i.e. with the quantum potential retained),  has been studied by Bilic {\it et al.} \cite{bilic} in relation with the formation of white dwarf stars by gravitational collapse.

The original Gross-Pitaevskii equation corresponds to a potential of the form
\begin{equation}
\label{eos5}
h(\rho)=g\rho=\frac{4\pi a_s\hbar^2}{m^3}\rho,
\end{equation}
where $a_s$ is the s-scattering length (in $d=3$). This leads to an equation of state
\begin{equation}
\label{eos6}
p=\frac{1}{2}g\rho^2=\frac{2\pi a_s\hbar^2}{m^3}\rho^{2}.
\end{equation}
This is a polyropic equation of state of the form (\ref{sp4}) with $n=1$, $\gamma=2$ and  $K=g/2={2\pi a_s\hbar^2}/{m^3}$. The GPP system (\ref{eos4})-(\ref{mfgp7})  with $n=1$ has been studied by B\"ohmer \& Harko \cite{bohmer} and Chavanis \cite{paper1,paper2,cosmobec} in the context of dark matter.

{\it Remark:} as we have previously indicated, for a Bose-Einstein condensate, the pressure can be negative. This is the case, in particular, for a BEC with quartic self-interaction ($p\propto \rho^2\propto |\psi|^4$) described by the equation of state (\ref{eos6}) when the scattering length $a_s$ is negative. In terrestrial BEC experiments, some atoms like $^7{\rm Li}$ have a negative scattering length \cite{revuebec,fedichev}. Similarly, the constant $T_{eff}$ appearing in Eq. (\ref{eos1}) is just an ``effective'' temperature since it arises from a particular form of self-interaction and has nothing to do with the thermodynamical temperature (which here is $T=0$). This effective temperature can be positive or negative.

\subsection{Time-independent dissipative GP equation}
\label{sec_tigp}

If we consider a wavefunction of the form
\begin{equation}
\label{tigp1}
\psi({\bf r},t)=A({\bf r})e^{-i\frac{E}{\xi\hbar}(1-e^{-\xi t})},
\end{equation}
we obtain the  time-independent dissipative GP equation
\begin{eqnarray}
\label{tigp2}
-\frac{\hbar^2}{2m}\Delta\psi({\bf r})+m(\Phi({\bf r})+h(\rho)+\Phi_{ext}({\bf r}))\psi({\bf r})=E\psi({\bf r}),\nonumber\\
\end{eqnarray}
where $\psi({\bf r})\equiv A({\bf r})$ is real and $\rho({\bf r})=Nm\psi^2({\bf r})$. Dividing Eq. (\ref{tigp2}) by $\psi({\bf r})$, we get
\begin{equation}
\label{tigp3}
m\Phi+mh(\rho)+m\Phi_{ext}({\bf r})-\frac{\hbar^2}{2m}\frac{\Delta \sqrt{\rho}}{\sqrt{\rho}}=E,
\end{equation}
or, equivalently,
\begin{equation}
\label{tigp4}
m\Phi+mh(\rho)+m\Phi_{ext}+Q=E.
\end{equation}
This relation can also be derived from the dissipative quantum Hamilton-Jacobi equation (\ref{mad7}) by setting $S=-(E/\xi)(1-e^{-\xi t})$. Alternatively, it can be deduced from the quantum barotropic Euler equation (\ref{mad10}) which is equivalent to the GP equation. The steady state of the quantum barotropic Euler equation (\ref{mad10}), obtained by taking $\partial_t=0$ and ${\bf u}={\bf 0}$, satisfies
\begin{equation}
\label{tigp5}
\nabla p+\rho\nabla\Phi+\rho\nabla \Phi_{ext}-\frac{\hbar^2\rho}{2m^2}\nabla \left (\frac{\Delta\sqrt{\rho}}{\sqrt{\rho}}\right )={\bf 0}.
\end{equation}
This generalizes the usual condition of hydrostatic equilibrium by incorporating the contribution of the quantum potential. Equation (\ref{tigp5}) describes the balance between the gravitational attraction, the repulsion due to the quantum potential (Heisenberg uncertainty principle) and the repulsion (for $a_s>0$) or the attraction (for $a_s<0$) due to the short-range interactions (scattering). This equation is equivalent to Eq. (\ref{tigp3}). Indeed, integrating Eq. (\ref{tigp5}) using Eq. (\ref{mad11}), we obtain Eq. (\ref{tigp3}) where the eigenenergy  $E$ appears as a constant of integration. On the other hand, combining Eq. (\ref{tigp5}) with the Poisson equation (\ref{mfgp7}), we obtain the fundamental equation of hydrostatic equilibrium with quantum effects
\begin{equation}
\label{tigp6}
-\nabla\cdot \left (\frac{\nabla p}{\rho}\right )+\frac{\hbar^2}{2m^2}\Delta \left (\frac{\Delta\sqrt{\rho}}{\sqrt{\rho}}\right )=S_d G\rho+\Delta\Phi_{ext}.
\end{equation}
For a harmonic potential, $\Delta\Phi_{ext}=d\omega_0^2$. When quantum effects are ignored (classical limit or Thomas-Fermi approximation), we recover the classical fundamental equation of hydrostatic equilibrium
\begin{equation}
\label{tigp6class}
-\nabla\cdot \left (\frac{\nabla p}{\rho}\right )=S_d G\rho+\Delta\Phi_{ext}.
\end{equation}
Finally, for an equation of state of the form (\ref{eos6}), Eq. (\ref{tigp6}) becomes
\begin{equation}
\label{tigp7}
-g\Delta\rho+\frac{\hbar^2}{2m^2}\Delta \left (\frac{\Delta\sqrt{\rho}}{\sqrt{\rho}}\right )=S_d G\rho+\Delta\Phi_{ext}.
\end{equation}
This equation has been studied in \cite{paper1,paper2} in $d=3$ when $\Phi_{ext}=0$. There are two important limits: (i) the non-interacting limit corresponding to $g=a_s=0$ (or more generally $p=0$) and (ii) the Thomas-Fermi (TF) limit obtained by neglecting the quantum potential $Q$.

\subsection{Equation for the evolution of the density contrast}
\label{sec_ev}

We can now generalize the results of the first part of the paper to the quantum BSP system (\ref{mad16})-(\ref{mad17}). In this paper, we assume $\Phi_{ext}=0$. The background flow (collapsing uniform sphere) studied in Sec. \ref{sec_hom} is not modified since the quantum potential enters (\ref{mad16}) only through its gradient. Then, working in the comoving frame and extending the calculations of Sec. \ref{sec_coll}, we find that Eqs. (\ref{c12})-(\ref{c13}) are replaced by
\begin{eqnarray}
\label{ev1}
\xi\frac{\partial\delta}{\partial t}=\frac{1}{a^2}\nabla\cdot \biggl \lbrack c_s^2\nabla \delta+(1+\delta)\nabla\phi\nonumber\\
-\frac{\hbar^2 (1+\delta)}{2m^2a^2}\nabla\left (\frac{\Delta\sqrt{1+\delta}}{\sqrt{1+\delta}}\right )\biggr\rbrack,
\end{eqnarray}
\begin{equation}
\label{ev2}
\Delta\phi=S_d G \rho_b \delta a^2,
\end{equation}
where $c_s^2=p'(\rho)=p'(\rho_b (1+\delta))$. If we consider small perturbations $\delta\ll 1$, $\phi\ll 1$ around the homogeneous flow $\delta=\phi=0$, we obtain the linearized equation
\begin{equation}
\label{ev3}
\xi\frac{\partial\delta}{\partial t}=\frac{c_s^2}{a^2}\Delta\delta+S_d G \rho_b \delta -\frac{\hbar^2}{4m^2a^4}\Delta^2\delta,
\end{equation}
where $c_s^2=p'(\rho_b)$. Expanding the solution in Fourier modes of the form $\delta({\bf x},t)=\delta_{\bf k}(t)e^{i{\bf k}\cdot {\bf x}}$, we obtain the equation
\begin{eqnarray}
\label{ev4}
\xi\dot\delta+\left (\frac{\hbar^2k^4}{4m^2a^4}+\frac{c_s^2k^2}{a^2}-S_d G\rho_b\right )\delta=0,
\end{eqnarray}
for the evolution of the density contrast in the linear regime.

\subsection{Jeans type analysis in a static frame}
\label{sec_jq}

Let us first make the Jeans swindle and assume that the above equation is valid in a static frame $a=1$. In that case, the solution of Eq. (\ref{ev4}) is $\delta\propto e^{-i\omega t}$ leading to the dispersion relation
\begin{equation}
\label{jq1}
i\xi\omega=\frac{\hbar^2k^4}{4m^2}+c_s^2k^2-S_dG\rho.
\end{equation}
The perturbation evolves like $e^{\gamma t}$ with an exponential rate $\gamma=-i\omega$. This relation, like the original quantum Smoluchowski equation (\ref{mad16}), clearly shows the competition between the attractive gravitational force, the (repulsive or attractive)  pressure arising from short-range interactions and the repulsive  quantum pressure arising from the Heisenberg uncertainty principle. For the equation of state (\ref{eos6}), the square of the velocity of sound is
\begin{equation}
\label{jq2}
c_s^2=g\rho=\frac{4\pi a_s\hbar^2\rho}{m^3},
\end{equation}
and the dispersion relation can be rewritten
\begin{equation}
\label{jq3}
i\xi\omega=\frac{\hbar^2k^4}{4m^2}+g \rho k^2-S_dG\rho.
\end{equation}

In the non-interacting case $g=a_s=0$, the pressure is zero ($p=0$). The particles interact via gravity and they experience the effect of the quantum potential. The dispersion relation reduces to
\begin{equation}
\label{jq6}
i\xi\omega=\frac{\hbar^2k^4}{4m^2}-S_dG\rho,
\end{equation}
leading to the quantum Jeans  wavenumber
\begin{equation}
\label{jq7}
k_Q=\left (\frac{4S_d G \rho m^2}{\hbar^2}\right )^{1/4}.
\end{equation}
This characteristic wavenumber arises due to the interplay between gravity and quantum effects (Heisenberg's uncertainty principle). For $k<k_Q$, the system is unstable and the perturbations grow with a growth rate $\gamma>0$; for $k>k_Q$ the system is stable and the perturbations decay with a damping rate $\gamma<0$. Note that the growth rate is maximum for $k=0$ (corresponding to an infinite wavelength $\lambda=2\pi/k\rightarrow +\infty$) and its value is $\gamma_{max}=S_dG\rho/\xi$.

In the Thomas-Fermi approximation in which the quantum potential can be neglected, the particles interact via gravity and they experience a pressure due to short-range interactions. The dispersion relation reduces to
\begin{equation}
\label{jq4}
i\xi\omega=g \rho k^2-S_dG\rho.
\end{equation}
This is the classical Jeans dispersion relation of Sec. \ref{sec_jeans}. For $g<0$, the system is always unstable. For $g>0$, the Jeans wavenumber is
\begin{equation}
\label{jq5}
k_J=\left (\frac{S_d G\rho}{c_s^2}\right )^{1/2}=\left (\frac{S_d G}{g}\right )^{1/2}.
\end{equation}
This  characteristic wavenumber  arises due to the interplay between gravity and repulsive scattering. We note that the Jeans wavenumber is independent on the density (in $d=3$, it can be written $k_J=({Gm^3}/{a_s\hbar^2})^{1/2}$). For $k<k_J$, the system is unstable and the perturbation grows with a growth rate $\gamma>0$; for $k>k_J$ the system is stable and the perturbation decays with a damping rate $\gamma<0$. The maximum growth rate corresponds to $k=0$ leading to $\gamma_{max}=S_d G\rho/\xi$.

In the non-gravitational case ($G=0$), the dispersion relation reduces to
\begin{equation}
\label{jq8}
i\xi\omega=\frac{\hbar^2k^4}{4m^2}+g\rho k^2.
\end{equation}
For $g>0$, the system is always stable and the perturbation decays with a damping rate $\gamma<0$. For $g<0$, the critical wavenumber is
\begin{equation}
\label{jq9}
k_0=\left (\frac{4m^2 |c_s^2|}{\hbar^2}\right )^{1/2}=\left (\frac{4m^2 |g|\rho}{\hbar^2}\right )^{1/2}.
\end{equation}
This  characteristic wavenumber arises due to the interplay between quantum pressure and attractive scattering (in $d=3$, $k_0=({16\pi |a_s|\rho}/{m})^{1/2}$).  For $k<k_0$, the system is unstable and the perturbation grows with a growth rate $\gamma>0$; for $k>k_0$ the system is stable and the perturbation decays with a damping rate $\gamma<0$. The growth rate is maximum for $k_*=(2|c_s^2| m^2/\hbar^2)^{1/2}=(2|g|\rho m^2/\hbar^2)^{1/2}=k_0/\sqrt{2}$ and its value is $\gamma_*=c_s^4 m^2/\xi\hbar^2=g^2\rho^2m^2/\xi\hbar^2$ (in $d=3$, $k_*=({8\pi |a_s|\rho}/{m})^{1/2}$ and $\gamma_*={16\pi^2 a_s^2\hbar^2\rho^2}/\xi {m^4}$).

\begin{figure}[!h]
\begin{center}
\includegraphics[clip,scale=0.3]{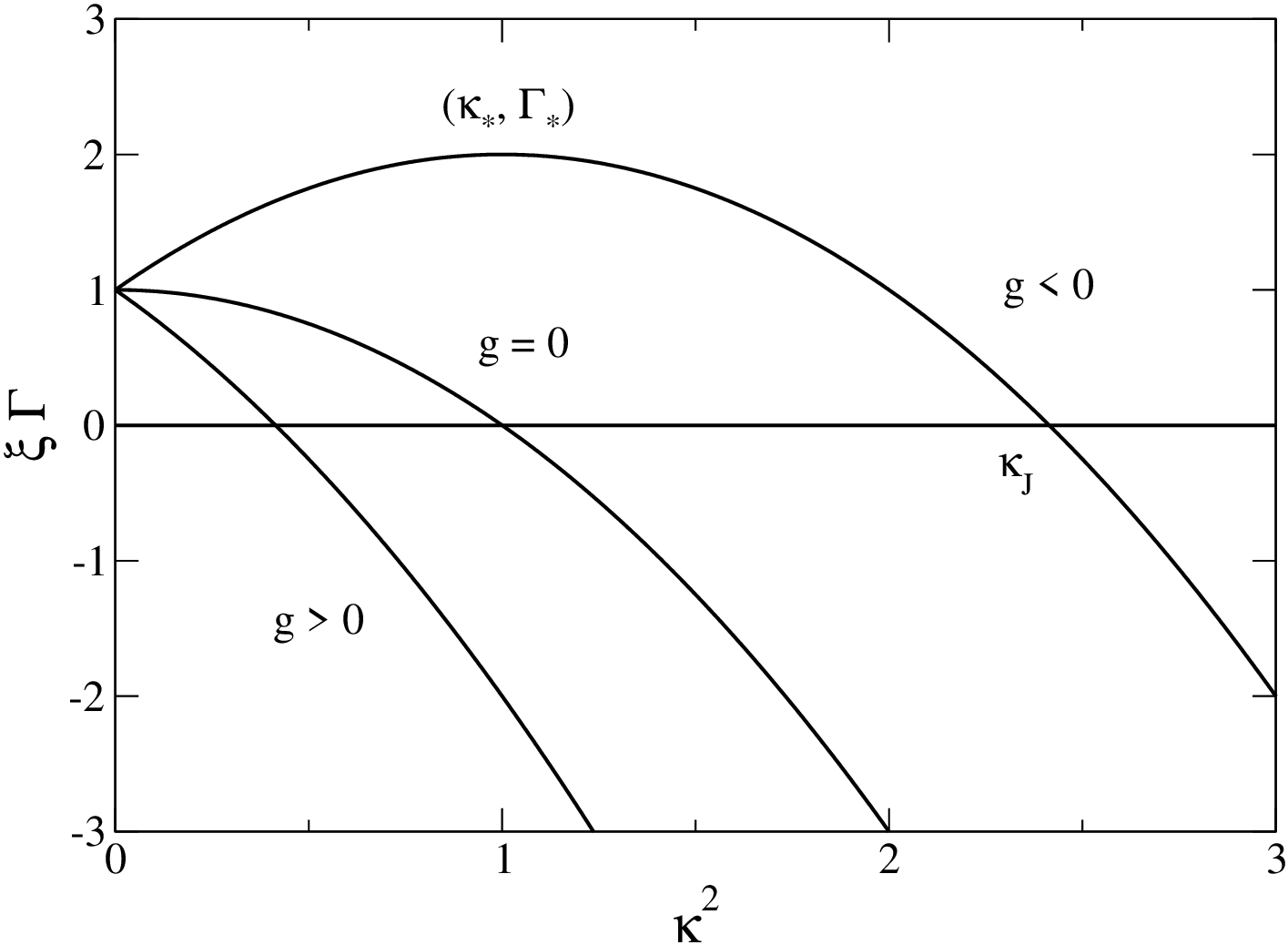}
\caption{Dimensionless dispersion relation $-\xi\Gamma=\kappa^4+2\alpha\kappa^2-1$ with $\Gamma=\gamma/\gamma_0$, $\kappa=k/k_0$ and $\alpha=g/g_0$ where $\gamma_0=S_d G\rho$, $k_0=(4S_d G\rho m^2/\hbar^2)^{1/4}$ and $g_0=(S_d G\hbar^2/\rho m^2)^{1/2}$. The Jeans wavenumber is $\kappa_J^2=-\alpha+\sqrt{\alpha^2+1}$. For $g<0$, the maximum growth rate is $\xi\Gamma_*={1+\alpha^2}$ reached for $\kappa_*^2=-\alpha$. The TF limit corresponds to $\kappa^2\ll |\alpha|$, the non-gravitational limit to $\kappa^2\gg 1/|\alpha|$ and the non-interacting limit to $|\alpha|\ll 1$.}
\label{exprate}
\end{center}
\end{figure}

We now consider the general case. With the previous relations, the dispersion relation (\ref{jq1}) can be rewritten
\begin{equation}
\label{jq9b}
\frac{i\xi\omega}{S_d G\rho}=\frac{k^4}{k_Q^4}+\frac{k^2}{k_J^2}-1.
\end{equation}
The pulsation vanishes at the critical Jeans wavenumber
\begin{equation}
\label{jq10}
k_c^2=\frac{k_Q^4}{2k_J^2}\left ( \pm\sqrt{1+\frac{4k_J^4}{k_Q^4}}-1\right ),
\end{equation}
with $+$ when $c_s^2\ge 0$ and $-$ when $c_s^2<0$. The previous results can be recovered as particular cases of this general relation. For $k<k_c$, the system is unstable and the perturbation grows with a growth rate $\gamma>0$; for $k>k_c$ the system is stable and the perturbation decays with a damping rate $\gamma<0$. For $c_s^2\ge 0$, $\gamma$ decreases monotonically with $k$. Accordingly, the growth rate is maximum for $k=0$ (infinite wavelength) leading to $\gamma_{max}=S_d G\rho/\xi$. For $c_s^2<0$, $\gamma$ achieves a maximum value
\begin{eqnarray}
\label{jq11}
\gamma_*=\frac{S_d G\rho}{\xi}\left (1+\frac{k_Q^4}{4k_J^4}\right )=\frac{S_d G\rho}{\xi}+\frac{m^2c_s^4}{\xi\hbar^2},
\end{eqnarray}
at
\begin{eqnarray}
\label{jq12}
k_*=\left (\frac{k_Q^4}{2|k_J^2|}\right )^{1/2}=\left (\frac{2|c_s^2|m^2}{\hbar^2}\right )^{1/2}.
\end{eqnarray}
The exponential rate $\gamma=-i\omega$ is plotted as a function of the wavenumber $k$ in Fig. \ref{exprate} for repulsive ($g>0$) and attractive ($g<0$) self-interactions.

{\it Remark:} if we perform the Jeans stability analysis on the quantum damped barotropic Euler-Poisson system (\ref{mad12})-(\ref{mad14}), we find that the dispersion relation is given by Eq. (\ref{jq1}) where $i\xi\omega$  is replaced by $\omega(\omega+i\xi)$. The discussion is then similar to the one given in Ref. \cite{chemojeans}.

\subsection{Jeans type analysis in a collapsing frame}
\label{sec_jqc}

We now return to Eq. (\ref{ev4}) and study the growth of perturbations in the collapsing frame. Taking  $c_s=0$ in Eq. (\ref{ev4}), we can define a quantum time-dependent Jeans wavenumber
\begin{eqnarray}
\label{jqc1}
k_Q=\left (\frac{4S_d G\rho_b m^2a^4}{\hbar^2}\right )^{1/4}.
\end{eqnarray}
Recalling Eq. (\ref{h3}), we can write $k_Q=\kappa_Q a^{(4-d)/4}$ with $\kappa_Q=({4 S_d G\rho_b a^d m^2}/{\hbar^2})^{1/4}$. The quantum Jeans length $\lambda_Q=2\pi/k_J$ behaves like $a^{-(4-d)/4}$. For $d>4$ (resp. $d<4$), it decreases (resp. increases) with time so that the system becomes unstable at smaller and smaller (larger and larger) scales as the cloud collapses. For $d=4$, the quantum Jeans length is constant: $k_J=\kappa_J$.

Measuring the evolution in terms of $a$ instead of $t$ and using Eq. (\ref{h8}), we can rewrite Eq. (\ref{ev4}) in the form
\begin{eqnarray}
\label{jqc2}
\frac{d\delta}{da}-\frac{d}{a}\left (\frac{\hbar^2k^4}{4S_d G\rho_b m^2a^4}+\frac{c_s^2k^2}{S_d G\rho_b a^2}-1\right )\delta=0.
\end{eqnarray}
For short times, writing $a=a_0(1-\epsilon)$ with $\epsilon\ll 1$, we obtain
\begin{eqnarray}
\label{jqc3}
\frac{\delta (a)}{\delta (a_0)}\simeq 1-d\left (\frac{k^4}{k_Q(0)^4}+\frac{k^2}{k_J(0)^2}-1\right )\epsilon,
\end{eqnarray}
where $k_Q(0)$ and $k_J(0)$ are the quantum and classical Jeans wavenumbers (\ref{jqc1}) and (\ref{lin3}) at $t=0$. For short times, the perturbations start to grow if $k<k_c(0)$ and start to decay if $k>k_c(0)$ where $k_c$ is defined by Eq. (\ref{jq10}). This result is valid for any equation of state. To describe larger times, it is necessary to specify the equation of state. Assuming that $p(\rho)$ is given by a polytropic equation of state and introducing the notations defined previously, Eq. (\ref{jqc2}) becomes
\begin{eqnarray}
\label{jqc4}
\frac{d\delta}{da}-\frac{d}{a}\left (\frac{k^4}{\kappa_Q^4 a^{4-d}}+\frac{k^2}{\kappa_J^2a^{d\gamma-2(d-1)}}-1\right )\delta=0.
\end{eqnarray}

Let us first consider the non-interacting case $c_s=0$. Eq. (\ref{jqc4}) reduces to
\begin{eqnarray}
\label{jqc5}
\frac{d\delta}{da}-\frac{d}{a}\left (\frac{k^4}{\kappa_Q^4 a^{4-d}}-1\right )\delta=0.
\end{eqnarray}
If $d\neq 4$, the solution of this equation is
\begin{eqnarray}
\label{jqc6}
\delta (a)\propto \frac{1}{a^d}e^{-\frac{dk^4}{\kappa_Q^4 (4-d)a^{4-d}}}.
\end{eqnarray}
For short times,  we recover Eq. (\ref{jqc3}). This equation clearly shows the initial effect of the Jeans wavelength. Indeed, the perturbation starts to grow if $k<k_Q(0)$ and starts to decay if $k>k_Q(0)$. This result is independent on the value of the dimension $d$. On the other hand, for large times, the effect of the quantum Jeans scale disappears and the evolution of the system is only controlled by the value of $d$. For $d<4$, we find that $\delta(a)\rightarrow 0$ very rapidly as $a\rightarrow 0$. The system becomes asymptotically stable to all wavelengths. For $d>4$, we find that $\delta(a)\propto a^{-d}\rightarrow +\infty$ as $a\rightarrow 0$. The system becomes asymptotically unstable to all wavelengths and behaves like in a cold classical gas. We note that quantum mechanics has a stabilizing role with respect to gravitational collapse when $d<4$. However, it is not able to prevent gravitational collapse when $d>4$. Therefore, the dimension $d=4$ is critical. We have already reached this conclusion in \cite{fermid,wdsd} for fermion stars.  For $d=4$, the solution of Eq. (\ref{jqc5}) is
\begin{eqnarray}
\label{jqc7}
\delta (a)\propto a^{4\left (k^4/\kappa_Q^4-1\right )}.
\end{eqnarray}
For $k<\kappa_Q$, the perturbation increases in time and the system is unstable (for $k\ll \kappa_Q$, the perturbation grows like in a cold classical gas). For $k>\kappa_Q$, the perturbation decreases in time and the system is stable.

Comparing  Eqs. (\ref{jqc5}) and (\ref{sol4}), we note that the dependence in $a$ is the same in the two equations when $\gamma=\gamma_{5/3}\equiv (d+2)/d$, i.e. $n=n_{3/2}\equiv d/2$. This is the index of a classical fermionic gas \cite{fermid,wdsd}.  Concerning the dependence in $a$, a cold bosonic gas behaves similarly to a cold fermionic gas. However, the dependence in $k$ is different ($k^2$ instead of $k$) as well as the expression of the Jeans length ($k_Q$ instead of $k_J$). We also note that $\gamma_{5/3}=\gamma_{4/3}=3/2$ for $d=4$, which is the critical dimension of a self-gravitating quantum gas \cite{fermid,wdsd}. For $\gamma=\gamma_{5/3}$, Eq. (\ref{jqc4}) reduces to
\begin{eqnarray}
\label{jqc8}
\frac{d\delta}{da}-\frac{d}{a}\left (\frac{k^4}{\kappa_Q^4 a^{4-d}}+\frac{k^2}{\kappa_J^2a^{4-d}}-1\right )\delta=0.
\end{eqnarray}
If $d\neq 4$, the solution of this equation is
\begin{eqnarray}
\label{jqc9}
\delta (a)\propto \frac{1}{a^d}e^{-\frac{d(k^4/\kappa_Q^4+k^2/\kappa_J^2)}{ (4-d)a^{4-d}}}.
\end{eqnarray}
For short times, we recover Eq. (\ref{jqc3}). The perturbation starts to grow if $k<k_c(0)$ and starts to decay if $k>k_c(0)$. For large times, the system is asymptotically stable for $d<4$ and asymptotically unstable for $d>4$. For $d=4$, the solution of Eq. (\ref{jqc8}) is
\begin{eqnarray}
\label{jqc10}
\delta (a)\propto a^{4\left (\frac{k^4}{\kappa_Q^4}+\frac{k^2}{\kappa_J^2}-1\right )}.
\end{eqnarray}
The system is stable for $k>k_c$ and unstable for $k<k_c$.

Let us finally treat the general case. We first assume $d\neq 4$. For $\gamma\neq \gamma_{4/3}$, we get
\begin{eqnarray}
\label{jqc11}
\delta (a)\propto \frac{1}{a^d}e^{-\frac{dk^4}{\kappa_Q^4 (4-d)a^{4-d}}}e^{\frac{dk^2a^{2(d-1)-d\gamma}}{\kappa_J^2 (2(d-1)-d\gamma)}},
\end{eqnarray}
and for $\gamma=\gamma_{4/3}$, we obtain
\begin{eqnarray}
\label{jqc12}
\delta (a)\propto \frac{1}{a^d}e^{-\frac{dk^4}{\kappa_Q^4 (4-d)a^{4-d}}}a^{d\left (k^2/\kappa_J^2-1\right )}.
\end{eqnarray}
We now assume $d=4$. For $\gamma\neq \gamma_{4/3}=3/2$, we get
\begin{eqnarray}
\label{jqc13}
\delta (a)\propto a^{4\left (k^4/\kappa_Q^4-1\right )} e^{\frac{4 k^2 a^{6-4\gamma}}{\kappa_J^2 (6-4\gamma)}},
\end{eqnarray}
and for $\gamma=\gamma_{4/3}=3/2$, we obtain Eq. (\ref{jqc10}).

The short time evolution ($a\rightarrow a_0$) is given by Eq. (\ref{jqc3}). The perturbation starts to grow if $k<k_c(0)$ and starts to decay if $k>k_c(0)$. To analyze the evolution for large time ($a\rightarrow 0$), it is convenient to start directly from the differential equation (\ref{jqc4}). (i) Let us first assume $d<4$. For $\gamma>\gamma_{5/3}$, $\delta(a)\rightarrow 0$  like in Eq. (\ref{sol5}) and for $\gamma<\gamma_{5/3}$, $\delta(a)\rightarrow 0$ like in Eq. (\ref{jqc6}). For $\gamma=\gamma_{5/3}$, $\delta(a)\rightarrow 0$ like in Eq. (\ref{jqc9}).
In all cases, we find that $\delta\rightarrow 0$ very rapidly so that the system is always asymptotically stable. The stabilization is due  to the polytropic pressure for $\gamma>\gamma_{5/3}$, to the quantum pressure for $\gamma<\gamma_{5/3}$ and to both for $\gamma=\gamma_{5/3}$. (ii) Let us consider $d>4$. In that case, the quantum effects become asymptotically negligible and we are led back to the study of Sec. \ref{sec_sol}. For $\gamma\neq \gamma_{4/3}$, the asymptotic evolution of the density contrast is given by Eq. (\ref{sol5}). For $\gamma>\gamma_{4/3}$, $\delta(a)\rightarrow 0$ very rapidly and the system is stable; for $\gamma<\gamma_{4/3}$, $\delta(a)\propto a^{-d}\rightarrow +\infty$ and the system is unstable. For $\gamma=\gamma_{4/3}$, the evolution of the density contrast is given by Eq. (\ref{sol8}). The system is stable for $k>k_J$  and unstable for $k<k_J$. (iii) Let us finally assume $d=4$. For $\gamma>3/2$, we find that $\delta(a)\rightarrow 0$ like in Eq. (\ref{sol5}) so the system is stable.  The quantum pressure is asymptotically negligible so the stabilization is due to the polytropic pressure. For $\gamma<3/2$, the polytropic pressure is asymptotically negligible. We find that $\delta(a)$ behaves like in Eq. (\ref{jqc7}) so that the system is stable for $k>k_Q$ and unstable for $k<k_Q$. The stabilization is due to the quantum pressure. Finally, for $\gamma=3/2$, the density contrast  behaves like in Eq. (\ref{jqc10}) so that the system is stable for $k>k_c$ and unstable for $k<k_c$. The stabilization is due to the quantum pressure and to the polytropic pressure.

{\it Remark 1:} a standard self-gravitating BEC with $\gamma=2$ is always stable since $2>\gamma_{4/3}$ in any dimension of space. In that case, $\kappa_J=(S_d G/g)^{1/2}$.

{\it Remark 2:} the previous discussion assumes that $c_s^2>0$. If the self-interaction is attractive ($c_s^2<0$, $\kappa_J^2<0$), the system becomes unstable in the following cases: (i) if $d<4$ and $\gamma>\gamma_{5/3}$; (ii) if $d<4$ and $\gamma=\gamma_{5/3}$ and $k<(\kappa_Q^4/|\kappa_J^2|)^{1/2}$; (iii) if $d>4$; (iv) if $d=4$ and $\gamma>3/2$. The other results are unchanged.

\section{Conclusion}
\label{sec_conclusion}

In this paper, we have studied the dynamical evolution of a collapsing
cloud of classical and quantum self-gravitating Brownian
particles.   A spatially homogeneous sphere is unstable and
collapses under its own gravity. This leads to a finite-time
singularity in which the system has a vanishing radius and an infinite
density. However, the collapsing sphere (background flow) may be unstable
to small perturbations and a process of fragmentation takes place.
This leads to the formation of dense localized clusters. We have
derived an exact set of equations (\ref{c12})-(\ref{c13}) describing
the evolution of the density contrast in the comoving frame.  In the
linear regime, the evolution of the perturbations can be described
analytically for any equation of state. Specific attention has been
given to the polytropic equation of state for the sake of
illustration. If $\gamma>\gamma_{4/3}$, the system is asymptotically
stable. However, the perturbations initially grow if their wavelength
$\lambda$ is larger than the Jeans scale $\lambda_J(0)$. This initial
growth can trigger nonlinear effects and result first in a cold
collapse then in the formation of dense localized clusters in virial
equilibrium when pressure effects become important. They correspond to
stable steady states of the BSP system (\ref{sp1})-(\ref{sp2}) in
which gravitational forces are balanced by pressure forces. If
$\gamma<\gamma_{4/3}$, the system is unstable. Small perturbations
grow in the linear regime whatever their wavelength. In the nonlinear
regime, these perturbations amplify and form dense localized
clusters. Their structure and evolution can be described by the BSP
system (\ref{sp1})-(\ref{sp2}), neglecting locally the contraction of
the background flow. These clusters are unstable and they experience
gravitational collapse. They can break into smaller fragments which
themselves break into smaller fragments in a hierarchical
manner. Alternatively, they can undergo a spherically symmetric
self-similar collapse up to a Dirac peak. At large scales, we can
ignore their internal structure and we are led to a reduced dynamical
system of ${\cal N}$ clusters in interaction. The clusters merge when
they come into contact and collapse on each other. Therefore, their
mass increases while their number decreases. This is reminiscent of a
coarsening process in statistical mechanics or in 2D decaying
turbulence. We have given preliminary analytical results valid in the
nonlinear regime. A more detailed description of the nonlinear regime
requires numerical simulations. This is a problem left for future
works. Throughout our study, we have found interesting
analogies with cosmology (in a universe experiencing a ``big-crunch'')
\cite{peebles,paddycosmo} and with stellar formation (in a collapsing molecular
cloud) \cite{hoyle,hunter1,hunter}. We have shown that the methods
initially devised in astrophysics (Euler-Poisson) could be extended to
a different context (Smoluchowski-Poisson) leading to interesting new
results and applications. We have also stressed the specificities of
our approach and the differences with the astrophysical studies.
Since the equations (\ref{c12})-(\ref{c13}) that we have derived for
self-gravitating Brownian particles are simpler than those used in
cosmology and astrophysics (because inertial effects are neglected in
our model), these equations could be used as a valuable prototype to
test the theories that have been developed in other domains to study
the process of self-organization in complex media.  We hope to develop
these issues in future works.

\appendix

\section{Repulsive interactions}
\label{sec_rep}

It can be of interest to consider the case of Brownian particles with
repulsive interactions (see Ref. \cite{zero} for further discussions
and applications). The basic equations of the paper remain valid
provided that $G$ is replaced by $-G$. Considering a spatially
homogeneous solution of the BSP system, as in Sec. \ref{sec_hom}, we
now find that the sphere expands, instead of contracting, as described
in Ref. \cite{zero}. Then, from the results of Sec. \ref{sec_lin}
properly adapted to this new situation (i.e. making the substitution
$G\rightarrow -G$), we find that the homogeneous solution is always
stable since $\dot\delta<0$ according to Eq. (\ref{lin2}). This is
consistent with the observation made in Ref. \cite{zero} that the
homogeneous expanding sphere is an ``attractor'' of the $T=0$ dynamics
for large times. Indeed, it is found in Appendix E of \cite{zero} that
the inhomogeneous density profiles $\rho({r},t)$ converge towards the
homogeneous expanding sphere for $t\rightarrow +\infty$.

\section{Cosmological constant}
\label{sec_cc}

If we introduce a ``cosmological constant'' $\Lambda$ in the Poisson equation (this terminology is abusive here since we are not dealing with cosmology) and write $\Delta\Phi=S_dG\rho-\Lambda$, the equation (\ref{h8}) for the scale factor $a(t)$ becomes
\begin{equation}
\label{cc1}
\xi\frac{da}{dt}=-\frac{GM}{a^{d-1}}+\frac{1}{d}\Lambda a.
\end{equation}
A {\it static} uniform sphere of density $\rho_b=\Lambda/S_d G$ and radius $a=(dGM/\Lambda)^{1/d}$ is a particular solution of the equations. To investigate its stability, we set $a(t)=a+\delta a(t)$ where $\delta a(t)$ is a small perturbation. Substituting this decomposition in Eq. (\ref{cc1}) and keeping only terms that are linear in $\delta a(t)$, we get
\begin{equation}
\label{cc2}
\xi\frac{d\delta a}{dt}=\Lambda \delta a,
\end{equation}
so that the perturbation grows as $\delta a\propto e^{\Lambda t/\xi}$. Therefore, the static solution is strongly unstable similarly to the Einstein static universe in cosmology.

\section{Fourier transform of the equations of motion}
\label{sec_der}

In this Appendix, we derive an exact equation for the Fourier transform of the discrete density contrast $\delta_{\bf k}^{(d)}(t)$ directly from the equations of motion of the overdamped Brownian particles at $T=0$. This derivation closely follows the derivation given by Peebles \cite{peebles} and Padmanabhan \cite{paddycosmo} in cosmology for particles with inertia. At $T=0$, the equations of motion are
\begin{equation}
\label{der1}
\xi\frac{d{\bf r}_i}{dt}=-\nabla_i\Phi_d,\qquad \Delta\Phi_d=4\pi G\rho_d,
\end{equation}
where $\rho_d({\bf r},t)=m\sum_i\delta_D\lbrack {\bf r}-{\bf r}_i(t)\rbrack$ is the discrete density field expressed as a sum of Dirac distributions and $\Phi_d({\bf r},t)$ is the corresponding gravitational potential. Making the changes of variables (\ref{c1}) and (\ref{c4}), we find that the equations of motion in the comoving frame are
\begin{equation}
\label{der2}
\xi\frac{d{\bf x}_i}{dt}=-\frac{1}{a^2}\nabla_i\phi_d,\qquad \Delta\phi_d=S_d G\rho_b a^2\delta_d.
\end{equation}
The discrete density field can be written
\begin{eqnarray}
\label{der4}
1+\delta_d({\bf x},t)\equiv \frac{\rho_d({\bf x},t)}{\rho_b}=\frac{V}{N}\sum_i\delta_D\lbrack {\bf x}-{\bf x}_i(t)\rbrack\nonumber\\
=\int \delta_D\lbrack {\bf x}-{\bf x}_T(t,{\bf q})\rbrack\, d{\bf q},
\end{eqnarray}
where ${\bf x}_T(t,{\bf q})$ is the (Lagrangian) position at time $t$ of the particle initially located at ${\bf x}={\bf q}$  (the subscript $T$ stands for ``trajectory''). Using Eq. (\ref{der4}), the Fourier transform of the density contrast is given by
\begin{eqnarray}
\label{der5}
\delta^{(d)}_{\bf k}(t)=\int e^{-i{\bf k}\cdot {\bf x}_T(t,{\bf q})}\, d{\bf q}-(2\pi)^d\delta_D({\bf k}).
\end{eqnarray}
Taking the time derivative of this expression and using Eqs. (\ref{der2}) and  (\ref{eie5}), we find that the equation satisfied by  $\delta^{(d)}_{\bf k}(t)$ is
\begin{eqnarray}
\label{der6}
\xi{\dot \delta}_{\bf k}^{(d)}=S_d G\rho_b\delta_{\bf k}^{(d)}
+\frac{S_dG\rho_b}{2}\int \delta_{{\bf k}'}^{(d)}\delta_{{\bf k}-{\bf k}'}^{(d)}\nonumber\\
\times\left\lbrack \frac{{\bf k}\cdot {\bf k}'}{{k'}^2}+\frac{{\bf k}\cdot ({\bf k}-{\bf k}')}{|{\bf k}-{\bf k'}|^2} \right\rbrack \, \frac{d{\bf k}'}{(2\pi)^d}.
\end{eqnarray}
This equation has the same form as Eq. (\ref{eie7}) with $T=0$. However, its interpretation is different because, in the present derivation, we have not made the mean field approximation. Therefore, Eq. (\ref{der6}) is exact and contains the same information as Eq. (\ref{der2}). It is valid for the Fourier transform of the {\it discrete} density contrast $\delta_d({\bf x},t)$ expressed in terms of Dirac distributions while Eq. (\ref{eie7}) is valid for the Fourier transform of the {\it smooth} (locally averaged) density contrast $\delta({\bf x},t)$. Although exact, Eq. (\ref{der6}) is not very useful in practice since it is difficult to deal with Dirac distributions.  Indeed, it is easier to directly integrate the equations of motion (\ref{der2}) that are equivalent to Eq. (\ref{der6}). By contrast,  Eq. (\ref{eie7}) relies on a mean field approximation but it applies to  a smooth density field which is more physically relevant.

The discrete density $\rho_{d}({\bf r},t)$ and the discrete density contrast $\delta_d({\bf x},t)$ satisfy equations similar to Eqs. (\ref{sp1}) and (\ref{c12}) with $T=0$ constructed with the discrete gravitational potentials  $\Phi_{d}({\bf r},t)$ and $\phi_d({\bf x},t)$. Although these equations are formally similar, their interpretation is different as explained previously. This is the same distinction as between the Klimontovich (exact) and the Vlasov (mean field) equations in plasma physics and stellar dynamics. More details can be found in Ref. \cite{hb5}.

Finally, it is possible to include stochastic forces (when $T\neq 0$) in the equations of motion. In the inertial frame, Eq. (\ref{der1}) is replaced by
\begin{equation}
\label{der7}
\frac{d{\bf r}_i}{dt}=-\frac{1}{\xi}\nabla_i\Phi_d+\sqrt{\frac{2k_B T}{\xi m}}{\bf R}_{i}(t),
\end{equation}
where ${\bf R}_{i}(t)$ is a white noise satisfying $\langle {\bf R}_i(t)\rangle ={\bf 0}$ and $\langle R_i^\alpha(t)R_j^\beta(t')\rangle=\delta_{ij}\delta_{\alpha\beta}\delta(t-t')$. In the comoving frame, Eq. (\ref{der2}) is replaced by
\begin{equation}
\label{der8}
\frac{d{\bf x}_i}{dt}=-\frac{1}{\xi a^2}\nabla_i\phi_d+\sqrt{\frac{2k_B T}{\xi m a^2}}{\bf R}_{i}(t).
\end{equation}
Following Ref. \cite{hb5}, it is possible to  derive exact equations satisfied by the discrete density field and by the discrete density contrast. When a mean field approximation is implemented, we find that the average density field and the average density contrast satisfy the equations (\ref{sp1})-(\ref{sp2}) and (\ref{c12})-(\ref{c13}) derived in the main text (in the isothermal case).

\section{Self-similar solution in physical space}
\label{sec_sss}

At $T=0$, the equation for the density contrast is
\begin{equation}
\label{sss1}
\xi\frac{\partial\delta}{\partial t}=S_d G\rho_b \delta (1+\delta)+\frac{1}{a^2}\nabla\delta\cdot \nabla\phi,
\end{equation}
\begin{equation}
\label{sss2}
\Delta\phi=S_d G \rho_b \delta a^2.
\end{equation}
Measuring the evolution in terms of $a$ rather than in terms of $t$, and using Eq. (\ref{h8}), we can rewrite Eq. (\ref{sss1}) as
\begin{equation}
\label{sss3}
\frac{a}{d}\frac{\partial\delta}{\partial a}+\delta+\delta^2=-\frac{1}{S_d G\rho_ba^2}\nabla\delta\cdot \nabla\phi.
\end{equation}
We consider a solution of the form
\begin{eqnarray}
\label{sss4}
\delta({\bf x},a)=\delta(a)D({\bf x}),\qquad \phi({\bf x},a)=F(a)S_d G\rho_b a^d Q({\bf x}),\nonumber\\
\end{eqnarray}
with
\begin{equation}
\label{sss5}
F(a)=a^{2-d}\delta(a),\qquad \Delta Q=D({\bf x}).
\end{equation}
Substituting this ansatz in Eq. (\ref{sss3}), we obtain the separate equations
\begin{equation}
\label{sss6}
\frac{a}{d}\frac{d\delta}{da}+\delta(a)=\mu\delta^2(a),
\end{equation}
\begin{equation}
\label{sss7}
\mu D({\bf x})=-D^2({\bf x})-\nabla D\cdot \nabla Q.
\end{equation}
These equations are equivalent to Eqs. (\ref{eie20}) and (\ref{eie21}). The equation for $\delta(a)$ can be solved as in Sec. \ref{sec_sel}. On the other hand, combining Eq. (\ref{sss7}) with the Poisson equation (\ref{sss5})-b, we find that $D({\bf x})$ satisfies the nonlinear differential equation
\begin{equation}
\label{sss8}
2\nabla\cdot \left (\frac{\nabla D}{\Delta\ln D}\right )+D=0.
\end{equation}

\section{$H$-theorem}
\label{sec_ht}

The free energy associated with the dissipative GPP system (\ref{mfgp6})-(\ref{mfgp7}), or equivalently with the dissipative quantum barotropic Euler-Poisson system (\ref{mad12})-(\ref{mad14}),  can be written
\begin{eqnarray}
\label{ef1}
F=\Theta_c+\Theta_Q+U+W+W_{ext}.
\end{eqnarray}
The first two terms correspond to the total kinetic energy
\begin{eqnarray}
\label{ef2}
\Theta=\frac{N\hbar^2}{2m}\int |\nabla\psi|^2 \, d{\bf r}.
\end{eqnarray}
Using the Madelung transformation, it can be decomposed into the  ``classical'' kinetic energy
\begin{eqnarray}
\label{ef3}
\Theta_c=\int\rho \frac{{\bf u}^2}{2}\, d{\bf r},
\end{eqnarray}
and the ``quantum'' kinetic energy
\begin{equation}
\label{ef4}
\Theta_Q=\frac{1}{m}\int \rho Q\, d{\bf r}.
\end{equation}
Substituting Eq. (\ref{mad8}) in Eq. (\ref{ef4}), we obtain the equivalent expressions
\begin{eqnarray}
\label{ef5}
\Theta_Q&=&-\frac{\hbar^2}{2m^2}\int \sqrt{\rho}\Delta\sqrt{\rho}\, d{\bf r}\nonumber\\
&=&\frac{\hbar^2}{2m^2}\int (\nabla\sqrt{\rho})^2\, d{\bf r}
=\frac{\hbar^2}{8m^2}\int \frac{(\nabla\rho)^2}{\rho}\, d{\bf r}.
\end{eqnarray}
The third term is the internal energy
\begin{eqnarray}
\label{ef6}
U&=&\int\rho\int^{\rho}\frac{p(\rho_1)}{\rho_1^2}\, d\rho_1\, d{\bf r}\nonumber\\
&=&\int \left\lbrack \rho h(\rho)-p(\rho)\right \rbrack\, d{\bf r}=\int H(\rho)\, d{\bf r},
\end{eqnarray}
where we have used Eq. (\ref{mad11}) and integrated by parts to obtain the second equality.
For a polytropic equation of state $p=K\rho^{\gamma}$, it takes the form
\begin{eqnarray}
\label{ef7}
U=\frac{K}{\gamma-1}\int \rho^{\gamma}\, d{\bf r}=\frac{1}{\gamma-1}\int p\, d{\bf r}.
\end{eqnarray}
In particular, for the potential (\ref{eos5}), using Eq. (\ref{eos6}), we get
\begin{eqnarray}
\label{ef8}
U=\frac{1}{2}g\int \rho^2\, d{\bf r}=\frac{2\pi a_s\hbar^2}{m^3}\int \rho^2\, d{\bf r}.
\end{eqnarray}
The fourth term is the gravitational potential energy
\begin{eqnarray}
\label{ef9}
W=\frac{1}{2}\int\rho\Phi\, d{\bf r}.
\end{eqnarray}
The fifth term is the external potential energy
\begin{eqnarray}
\label{ef10}
W_{ext}=\int\rho\Phi_{ext}\, d{\bf r}.
\end{eqnarray}
For a harmonic potential, $W_{ext}=(1/2)\omega_0^2 I$ where $I$ is the moment of inertia (\ref{virial1}). It is easy to establish \cite{paper1} that $\delta \Theta_c=\int  ({\bf u}^2/2)\delta\rho\, d{\bf r}+\int \rho {\bf u}\cdot \delta {\bf u}\, d{\bf r}$, $\delta \Theta_Q=(1/m)\int Q \delta\rho \, d{\bf r}$, $\delta U=\int h(\rho)\delta\rho\, d{\bf r}$, $\delta W=\int \Phi\delta\rho\, d{\bf r}$ and $\delta W_{ext}=\int \Phi_{ext}\delta\rho\, d{\bf r}$. Taking the time derivative of $F$, using the previous relations and the hydrodynamic equations (\ref{mad12})-(\ref{mad14}), we easily obtain the $H$-theorem
\begin{eqnarray}
\label{ef11}
\dot F=-\xi\int \rho {\bf u}^2\, d{\bf r}\le 0.
\end{eqnarray}
This result remains valid if ${\bf u}$ is not a potential flow since ${\bf u}\cdot ({\bf u}\times (\nabla\times {\bf u}))={0}$. For a steady state, $\dot F=0$.  Equation (\ref{ef11}) implies ${\bf u}={\bf 0}$ and we recover the condition of hydrostatic equilibrium (\ref{tigp5}).

The free energy associated with the quantum BSP system (\ref{mad16})-(\ref{mad17}) is
\begin{eqnarray}
\label{ef12}
F=\Theta_Q+U+W+W_{ext}.
\end{eqnarray}
Taking the time derivative of $F$ and using the previous relations, we easily obtain the $H$-theorem
\begin{eqnarray}
\label{ef13}
\dot F=-\int \frac{1}{\xi \rho}(\nabla p+\rho\nabla\Phi+\rho\nabla\Phi_{ext}+\frac{\rho}{m}\nabla Q)^2\, d{\bf r}\le 0.\nonumber\\
\end{eqnarray}
It can also be obtained from Eq. (\ref{ef11}), using Eq. (\ref{mad15}). For a steady state, $\dot F=0$. Equation (\ref{ef13})  implies that the term in parenthesis vanishes leading to the condition of hydrostatic equilibrium (\ref{tigp5}).

These results indicate that the system will converge towards a steady state of the dynamical equations that is a (local) {\it minimum} of free energy at fixed mass (the mass $M=\int \rho\, d{\bf r}$ is a conserved quantity). Maxima or saddle points of free energy are unstable. If several local minima of free energy exist, the selection will depend on a notion of basin of attraction. We are therefore led to considering the minimization problem
\begin{eqnarray}
\label{ef14}
\min_{\rho,{\bf u}} \left\lbrace F[\rho,{\bf u}]\quad |\quad M\right\rbrace.
\end{eqnarray}
An extremum of free energy at fixed mass is given by the variational principle $\delta F-\alpha\delta M=0$
where $\alpha$ is a Lagrange multiplier taking into account the mass constraint. This gives ${\bf u}={\bf 0}$ and the condition
\begin{eqnarray}
\label{ef14b}
m\Phi+m\Phi_{ext}+m h(\rho)-\frac{\hbar^2}{2m}\frac{\Delta\sqrt{\rho}}{\sqrt{\rho}}=m\alpha.
\end{eqnarray}
This equation is equivalent to the steady state equation (\ref{tigp3}) provided that we make the identification
\begin{eqnarray}
\label{ef15}
\alpha=E/m.
\end{eqnarray}
This shows that the Lagrange multiplier (chemical potential) in the constrained minimization problem associated with (\ref{ef14}) is equal to the eigenenergy $E$ by unit of mass. On the other hand, considering the second order variations of free energy, we find that the steady state is stable iff
\begin{eqnarray}
\label{ef16}
\delta^2 F\equiv \frac{1}{2}\int h'(\rho)(\delta\rho)^2\, d{\bf r}+\frac{1}{2}\int \delta\rho\delta\Phi\, d{\bf r}\nonumber\\
+\frac{\hbar^2}{8m^2}\int  \left \lbrack \nabla \left (\frac{\delta\rho}{\sqrt{\rho}}\right )\right\rbrack^2\, d{\bf r}+\frac{\hbar^2}{8m^2}\int \frac{\Delta\sqrt{\rho}}{\rho^{3/2}}(\delta\rho)^2\, d{\bf r}>0,\nonumber\\
\end{eqnarray}
for all perturbations that conserve mass: $\int \delta\rho\, d{\bf r}=0$.

{\it Remark:} for $\xi=0$, the functional $F$ becomes the total energy $E_{tot}$. According to Eq. (\ref{ef11}), the total energy $E_{tot}$ is conserved by the GPP (or quantum barotropic Euler-Poisson)  system. In that context, the minimization problem (\ref{ef14}) provides a condition of nonlinear dynamical stability for a steady state of the GPP (or quantum barotropic Euler-Poisson) system \cite{paper1}.

\section{Virial theorem}
\label{sec_virial}

In this Appendix, we establish the time-dependent virial theorem associated with the dissipative quantum barotropic Euler-Poisson system (\ref{mad12})-(\ref{mad14}). The moment of inertia is
\begin{eqnarray}
\label{virial1}
I=\int \rho r^2\, d{\bf r}.
\end{eqnarray}
Taking its time derivative and using the continuity equation (\ref{mad12}), we obtain
\begin{eqnarray}
\label{virial2}
\dot I=2\int \rho {\bf r}\cdot {\bf u}\, d{\bf r}.
\end{eqnarray}
With the aid of the continuity equation, the quantum barotropic Euler equation (\ref{mad13}) can be rewritten
\begin{eqnarray}
\label{virial3}
\frac{\partial}{\partial t}(\rho {\bf u})+\nabla(\rho {\bf u}\otimes {\bf u})\qquad\qquad\qquad\nonumber\\
=-\nabla p-\rho\nabla\Phi-\rho\nabla \Phi_{ext}-\frac{\rho}{m}\nabla Q-\xi\rho {\bf u}.
\end{eqnarray}
Taking the time derivative of Eq. (\ref{virial2}), substituting Eq. (\ref{virial3}), and using integrations by parts, we obtain the time-dependent virial theorem
\begin{eqnarray}
\label{virial4}
\frac{1}{2}\ddot I+\frac{1}{2}\xi\dot I=2(\Theta_c+\Theta_Q)+d\int p\, d{\bf r}+W_{ii}+V_{ext}.
\end{eqnarray}
For a polytropic equation of state $p=K\rho^\gamma$, using Eq. (\ref{ef7}), we have $\int p\, d{\bf r}=(\gamma-1)U$. To obtain Eq. (\ref{virial4}), we have used the following identities. First, it can be established after a few manipulations (using essentially integrations by part \cite{paper1}) that
\begin{equation}
\label{virial5}
-\int \frac{\rho}{m}{\bf r}\cdot \nabla Q\, d{\bf r}=2\Theta_Q.
\end{equation}
On the other hand, we have introduced the virial of the gravitational force
\begin{equation}
\label{virial6}
W_{ii}=-\int \rho {\bf r}\cdot\nabla\Phi\, d{\bf r}.
\end{equation}
It can be shown that $W_{ii}=(d-2)W$ if $d\neq 2$ and $W_{ii}=-GM^2/2$ if $d=2$ \cite{virial1}. Finally, we have introduced the virial of the external force
\begin{equation}
\label{virial7}
V_{ext}=-\int \rho {\bf r}\cdot\nabla\Phi_{ext}\, d{\bf r}.
\end{equation}
For a harmonic potential, $V_{ext}=-\omega_0^2 I=-2W_{ext}$. 

In the strong friction limit $\xi\rightarrow +\infty$ in which ${\bf u}=O(1/\xi)$, the time-dependent virial theorem takes the form
\begin{equation}
\label{virial8}
\frac{1}{2}\xi\dot I=2\Theta_Q+d\int p\, d{\bf r}+W_{ii}+V_{ext}.
\end{equation}
It can be directly obtained from the quantum BSP system (\ref{mad16})-(\ref{mad17}).

For a steady state, $\dot I=\ddot I=0$ and ${\bf u}={\bf 0}$, we obtain the equilibrium virial theorem
\begin{equation}
\label{virial9}
2\Theta_Q+d\int p\, d{\bf r}+W_{ii}+V_{ext}=0.
\end{equation}
On the other hand, the free energy reduces to $F=\Theta_Q+U+W+W_{ext}$. Finally, multiplying the steady state equation (\ref{tigp4}) by $\rho/m$ and integrating over the whole configuration, we obtain the general identity
\begin{equation}
\label{virial10}
2W+\int\rho h(\rho)\, d{\bf r}+W_{ext}+\Theta_Q= N E.
\end{equation}
For a polytropic equation of state, using Eqs. (\ref{eos3}) and (\ref{ef7}), we find that $\int \rho h(\rho)\, d{\bf r}=\gamma U$.

\end{document}